\documentclass[12pt,preprint]{aastex}
\newcommand{\be}{\begin{equation}}
\newcommand{\ee}{\end{equation}}

\newcommand{\sd}[1]{\left \{ #1\right \}_s}

\shorttitle{Evolution of star clusters}
\shortauthors{Trenti et al.}

\begin{document}


\title{Tidal disruption, global mass function and structural parameters evolution in star
clusters}


\author{Michele Trenti}
\affil{University of Colorado, CASA, Dept.  of Astrophysical \& 
Planetary Sciences, 389-UCB, Boulder, CO 80309, USA} 
\email{trenti@colorado.edu}
\author{Enrico Vesperini}
\affil{Department of Physics, Drexel University, Philadelphia, PA 19104, USA}
\author{Mario Pasquato}
\affil{Dipartimento di Fisica, Universit\`a di Pisa, Largo Bruno Pontecorvo 3, 56127 Pisa, Italy}


\begin{abstract}

  We present a unified picture for the evolution of star clusters on
  the two-body relaxation timescale. We use direct N-body simulations
  of star clusters in a galactic tidal field starting from different
  multi-mass King models, up to 10\% of primordial
  binaries and up to $N_{tot}=65536$ particles. An additional run also
  includes a central Intermediate Mass Black Hole. We find that for
  the broad range of initial conditions we have studied the stellar mass
  function of these systems presents a universal evolution which
  depends only on the fractional mass loss. The structure
  of the system, as measured by the core to half mass
  radius ratio, also evolves toward a
  universal state, which is set by the efficiency of heating on the
  visible population of stars induced by dynamical interactions in the
  core of the system. Interactions with dark remnants
  (white dwarfs, neutron stars and stellar mass black holes) are dominant over the heating
  induced by a moderate population of primordial binaries (3-5\%),
  especially under the assumption that most of the neutron stars and
  black holes are retained in the system. All our models without
  primordial binaries undergo a deep
  gravothermal collapse in the radial mass
  profile. However their projected light distribution can be well
  fitted by
  medium concentration King models (with parameter $W_0 \sim 8$), even
  though there tends to be an excess over the best fit for the
  innermost points of the surface brightness. This excess is consistent with a shallow
  cusp in the surface brightness ($\mu \sim R^{-\nu}$ with $\nu \sim
  0.4-0.7$), like it has  
  been observed for many globular clusters from high-resolution HST imaging.
  Generally fitting a King
  profile to derive the structural parameters yields to larger
  fluctuations in the core size than defining the core as the radius
  where the surface brightness is one half of its central value.
  Classification of core-collapsed globular clusters based
  on their surface brightness profile may thus fail in systems that
  appear to have already bounced back to lower concentrations,
  particularly if the angular resolution of the observations is
  limited and the core is not well resolved. 
\end{abstract}


%%%%%%%%%%%%%%%%%%%%%%%%%%%%%%%%%%%%
\section{Introduction}\label{sec:intro}
Until a few years ago, the standard picture of globular clusters
stellar population and dynamical evolution emerging from theoretical
and observational studies was yielding a consistent and well
understood framework. Globular clusters were thought to be 'simple
stellar population', composed of stars with the same age and chemical
composition. Globular cluster dynamical evolution had been thoroughly
investigated by a large number of numerical studies showing that
clusters surviving the early expansion triggered by primordial gas
expulsion and mass loss due to stellar evolution evolve, due mainly to
two-body relaxation, toward core collapse and higher central
concentrations while losing stars through the boundary set by the
tidal field of their host galaxy (see e.g. \citealt{hh}).

In the last few years, however, a wealth of observational data
mostly from exquisite space based observations have revealed the
actual complexity of globular cluster dynamics and stellar
populations. It is now clear that there is  
a close link and interplay between dynamical evolution and the stellar
content of clusters, the structure of clusters and the
abundances of exotic objects (such as blue stragglers, X-ray sources,
pulsars etc.) (see e.g. \citealt{bel06,shara06,hut06}).

Recently, a number of photometric and spectroscopic observations have
also shown evidence of the presence of multiple stellar populations
and challenged the commonly held view according to which globular
cluster are 'simple stellar populations'. It appears that essentially
any globular cluster that has high quality photometric and
spectroscopic data available presents evidence against a single
star-formation burst \citep{gratton2004,bedin2004,piotto05,piotto07,carretta09}.
The presence of multiple stellar populations has major implications
for the formation and the early evolution of globular clusters (see
e.g. \citealt{ercole08}).

Furthermore, our understanding of the late-time evolution of globular
clusters has been called into question. A recent observational study
\citet{dem07},  focused on the stellar mass function of Galactic
globular clusters, found that clusters with lower concentration index
of the surface brightness profile tend to have flatter stellar mass
functions. The observed trend is at odds with what is expected from
the standard dynamical evolution scenario according to which as
clusters evolve toward core collapse and higher concentrations they
lose mass and the preferential loss of low-mass stars flattens the
stellar mass function; dynamically older clusters should be more
centrally concentrated and have flatter stellar mass functions.
\citet{bau08} have suggested that strong primordial mass segregation
might be required to explain the flat stellar mass function of some of
the low-concentration cluster in the observed sample of \citet{dem07}.
However a strong primordial mass segregation might lead to the rapid
dissolution of the cluster due to stellar evolution mass-losses
\citep{vesperini09}.

High-resolution photometry of the center of globular clusters has also
revealed that the standard \citet{king66} models do not capture the
details of the surface brightness profile, which often shows the
presence of shallow cusps within a relatively large core
\citep{noyola06}. The shallow cusps contrast with the classic
expectation that a strong, isothermal cusp develops as a result of
core-collapse. The presence of a central IMBH is a possible
explanation for the presence of those cusps \citep{bau04,tre07b}, but
this interpretation appears in contrast with the lack of strong
evidence for IMBHs in globular clusters \citep[e.g. see][and
references therein]{gill08}. NGC2298, for example, is one of the
clusters showing a central  
excess in the surface brightness profile from the \citet{noyola06}
data, but a central IMBH is ruled out at high confidence level
\citep{pas09}.

The goal of this paper is to study in detail the relation between a
cluster structural properties and the dynamical phases of its
evolutionary path as well as the relation between the properties of a
cluster stellar mass function and its structure. We focus our
attention on the long-term evolution of star clusters driven by the
effects of two-body relaxation. We analyze our numerical models
following the procedures used in the observational studies of the
structure and stellar content of star clusters and by directly
comparing our results with observational data we aim at
clarifying their interpretation in the context of dynamic evolution of
relaxed stellar systems.

We have carried out a survey of multimass simulations of star clusters
evolving in a tidal field and starting from a broad range of different
initial conditions spanning different initial density profiles,
primordial binary fractions and initial stellar mass function.
 This
paper is organized as follows. In Section~\ref{sec:sim} we present our
sample of simulations. In Section~\ref{sec:defs} we introduce our
framework for analyzing simulations consistently with the
observational derivation of structural quantities. In
Section~\ref{sec:results} we presents the results of our analysis,
concluding in Section~\ref{sec:conc}.

\section{Numerical framework}\label{sec:sim}

\subsection{Past investigations}

Numerical investigations of the dynamics of globular clusters are
computationally very challenging because of the need to resolve a
large dynamic range in timescales, from a few hours typical of a tight
binary to the billion of years over which the system evolves due to
two-body relaxation. In addition, the computational complexity of a
run carried out for a constant number of relaxation times scales as
$N^3/\log(N)$, because the relaxation time $t_{rh}$ increases with the
number of particles N (e.g see \citealt{spi87}). The computational
resources required are further increased by the presence of a
significant number of binaries, which can be up to $50 \%$ of the mass
of the core \citep[e.g., see][]{rub97,alb01,pul03}, although some
clusters, like NGC 6397, might also be almost binary-free
\citep{davis08}. Hard binaries have an important dynamic effect on the
evolution of the core size \citep{mcm90,mcm91,ves94,heg06}. In the
past, numerical simulations have been performed either using
approximate algorithms such as Monte Carlo methods \citep[e.g.,
see][]{gao91,fre07} or direct N-body simulations with a modest number
of particles ($N \approx 10^3$ in \citealt{mcm90} and \citealt{heg92})
until a recent improvement of one order of magnitude ($N=16384$ -
\citealt{heg06,tre07a,tre07b,tre08,hurley07}). These investigations
highlighted two key results: (i) the presence of a significant
population of primordial binaries drives the evolution of the core
radius toward a value that is $5-10\%$ of the half mass radius and
(ii) the ratio of the core to half mass radius ($r_c/r_h$) does not
depend either on its initial value or on the primordial binary
fraction $f$, once a few relaxation times have passed and provided
that $f \gtrsim 0.1$. In addition, runs with a galactic tidal field
show the central concentration parameter $c$ decreases toward the end
of the simulation, due to the combined effect of binaries, that keep a
large core, and of the evaporation, that progressively reduces the
tidal radius \citep{tre07a}. However, the simulations of
\citet{tre07a} were obtained using simplified initial conditions with
equal mass particles and therefore cannot be directly applied to the
study of the mass function evolution or analyzed consistently with
observations. In this paper we extend our earlier work to include a
realistic mass spectrum, as discussed below.

\subsection{Our simulations}

The simulation framework that we use here has been extensively
described in \citet{tre07a}, \citet{gill08} and \citet{pas09}. In
short, we follow the evolution of star clusters using the direct
summation code NBODY6 \citep{aar03}, which guarantees an exact
treatment of the multiple interactions between stars by employing
special regularization techniques, without resorting to the
introduction of softening.

Our N-body simulations are carried out in natural (dimensionless)
units \citep{heg86}, in which:
%%%%%
\be
G=M_T=-4E_T=1,
\ee
%%%%%
where $M_T$ is the total mass of the system and $E_T$ the total energy
(potential plus kinetic). The corresponding unit of time is:
%%%%%
\be
t_d = \frac{(GM_T)^{5/2}}{(−4E_T)^{3/2}} = 1, 
\ee
%%%%%
which approximately corresponds to the orbital period at the half-mass
radius. In these units, the relaxation time can be written as:
%%%%%%%%%%%
\be \label{trel}
t_{rh} = \frac{0.138 N r_h^{3/2}}{log(0.11 N)},
\ee
%%%%%%%%%%%
where $r_h \approx 1$ is the radius at which the relaxation time is
defined, that is the half-mass radius. Equation~\ref{trel} can be
translated in physical units as follows \citep{djo93}:
%%%%
\be 
t_{rh} = \frac{8.9\times 10^5 \mathrm{yr}}{\log{(0.11N)}} \times
\frac{1 M_{\sun}}{\langle m_* \rangle}\times
\left(\frac{M_T}{M_{\sun}}\right)^{1/2} \times \left( \frac{r_h}{1
   \mathrm{pc}}\right)^{3/2}, 
\ee
%%%%:
where $\langle m_* \rangle$ is the average mass of a star in the
system (including dark remnants such as neutron stars).

The individual particle mass is drawn from an initial mass function
appropriate to study the late evolutionary stages of star clusters,
when stars are $\gtrsim 10$~Gyr old. As in \citet{gill08} we consider
either a \citet{salp} or a \citet{ms} initial mass function, that is:
\be \xi(m) \propto m^{\alpha}, \ee with $\alpha=-2.35$ and $m \in
[0.2:100] M_{\sun}$ for the Salpeter IMF, while for the Miller \&
Scalo IMF the power-law slope is the following: $\alpha =-1.25$ for
$m\in [0.2:1] M_{\sun}$, $\alpha =-2.0$ for $m\in [1:2] M_{\sun}$,
$\alpha =-2.3$ for $m\in [2:10] M_{\sun}$, and $\alpha =-3.3$ for
$m\in [10:100] M_{\sun}$. Before starting the run, an instantaneous
step of stellar evolution is taken to evolve the mass function to a
$0.8~\mathrm{M_{\sun}}$ turnoff using the \citet{hurley2000}
evolutionary tracks. In our standard models we have a 100\% retention
fraction of dark remnants. We also run two simulations with a 30\%
retention fraction of neutron stars and stellar mass black holes (see
Table~\ref{tab:sim}). In all simulations no kick velocities are given
to the remnants. Our idealized treatment of stellar evolution is based
on the approximation that most of the relevant stellar evolution
occurs on a timescale shorter than a relaxation time. It is thus
appropriate to study the evolution of old globular clusters (age of
$\sim 10$ Gyr). In fact, most of the impact of stellar evolution on
the dynamics of a star cluster happens within the first few hundred
Myr, when the most massive stars lose a significant fraction of mass
and consequently contribute to a global expansion of the system (e.g.
see \citealt{hurley07,mackey08}). Later in the life of a star cluster,
two-body relaxation tends to erase the memory of the initial density
profile and concentration (\citealt{tre07a,tre08}).

The initial distribution in the position-velocity phase space is that
of a King model with scaled central potential $W_0=3,5,7$. Our
standard models have a primordial binary ratio of $f=N_b/(N_s+N_b) $
between 0 and 0.1, with $N_s$ and $N_b$ being the number of singles
and binaries respectively. In addition, we also consider a run with a
central Intermediate Mass Black Hole (discussed in more detail in
\citealt{gill08}). Following \citet{tre07a}, our initial distribution
of the binaries' binding energies is flat in log scale in the range
from $\epsilon_{min}$ to $133\epsilon_{min}$, with
$\epsilon_{min}=\langle m(0) \rangle \sigma_c(0)^2$. Here
$\sigma_c(0)$ is the initial central velocity dispersion and $\langle
m(0) \rangle$ is the average stellar mass at $t=0$. For a typical
velocity dispersion $\sigma_c(0)=10~\mathrm{km/s}$, the semi-axes of
binaries in the initial conditions are smaller than $10$ AU. This
choice is motivated by the fact that wider and therefore softer
binaries, which are likely created in young star clusters, are
destroyed within a few dynamical times and thus would have no effect
on the late-time evolution of a star cluster (\citealt{heggie75}). The
stellar mass and type of binary members are drawn at random from our
initial mass function and thus binaries containing either one or two
dark remnants are possible.

The models considered in this paper are tidally limited. The Galactic
tidal field is treated as that of a point mass, and the tidal force
acting on each particle is computed using a linear approximation of
the field. The tidal cutoff radius of the simulation (which we
indicate as $\sd{r_t}$, where the curl-bracket with the suscript ``s''
is used to denote the definition of a simulation based quantity, see
Section~\ref{sec:defs}) is defined as \citep{spi87}:
%%%%%%%%%%
\be\label{eq:rt}
\sd{r_t}^3 = \frac{M_T}{3M_{Gal}}R_{Gal}^3,
\ee
%%%%%%%%%%
where $M_{Gal}$ is the galaxy mass, and $R_{Gal}$ is the distance of
the centre of the star cluster from the centre of the galaxy. We
assume $R_{Gal}$ is constant, i.e. the cluster describes a circular
orbit. The initial value of $\sd{r_t}$ is fixed to be self-consistent
with the King profile used (for example, the $W_0=7$ king model has
$\sd{r_t}=7.02$ in N-body units). Particles are removed from the
system when they reach a distance $2\sd{r_t}$ from the center of
the cluster (see \citealt{tre07a} for a full description of the tidal
field treatment).

We explore a large parameter space in the initial conditions varying
the initial King model concentration, the Roche lobe filling factor,
the initial mass function and the primordial binary fraction, as
summarized in Table~\ref{tab:sim}. The NBODY6 code has been run with
its Graphic Processing Unit extension on the NCSA Lincoln Cluster.
Each simulation was assigned to a computing node with 8 cores and two
Tesla C1060 GPUs. We measured a sustained computational performance
above 0.5 Tflops for a single node, that allowed us to simulate
systems with up to $N=65536$ (or $N=64$k in a compact notation where
$1\mathrm{k} \equiv 1024$) particles until either $t=8000$ was reached
(corresponding to several initial relaxation times --- $t>
16~t_{rh}(0)$) or at least $\sim$ 80\% of the initial mass is lost due
to evaporation of stars.

%%%%%%%%%%%%%%%%%%%%
\section{Structural parameters of globular clusters: definitions in
  simulations and observations}\label{sec:defs}

Both space-based photometry with HST and adaptive optics from the
ground have greatly improved our knowledge of Galactic globular
clusters. In fact, the angular resolution currently available is such
that it is possible to resolve faint individual main sequence stars at
the center of most Galactic globulars. One of the most striking
examples is the globular clusters NGC 2298, where HST photometry
reaches a completeness greater than $50\%$ for $0.2~\mathrm{M_{\sun}}$
at the center of the system \citep{dem_pul,pas09}.

A large fraction of globular clusters have been observed with HST
\citep[e.g. see][]{sara07}, thus acquiring detailed information on
their current stellar mass function and their central surface
brightness profiles. Here we discuss the definition and determination
of structural parameters in both simulations and observations, with
the goal to present a consistent analysis of our numerical
investigation. Typically, structural properties in numerical
simulations are based on a complete knowledge of the system with a
tri-dimensional mass-based approach (e.g. the 3D half-mass radius of
the system). Observations rely instead on projected two-dimensional
data, that are light based (e.g. the 2D half-light radius). In
addition, observations are affected by finite angular resolution,
limited signal-to-noise and field of view. To ensure self-consistency
between quantities in simulations and observations, we developed a
framework to ``observe'' a simulation snapshot, from which we obtain
``observed'' structural parameters. For reference we also provide
structural parameters using the usual theoretical/numerical
definitions. Quantities that derive from the ``observation'' of a
snapshot have no markings.

We adopt instead a curl-bracketing with an underscore ``s'' ($\{
\}_s$) to denote the definition of a quantity based on current
practice in numerical simulations (the ``s'' subscript stands for
simulation).

\subsection{Observation of a simulation snapshot}

The positions and velocities of all particles in our simulations are
saved every $10$ code time units (one time unit is about one dynamic
time; see \citealt{heg86}). We then proceed to construct a synthetic
observation by first ignoring all particles except main sequence
stars. As our simulations only have an instantaneous step of stellar
evolution, we don't consider stars along the giant branch. This choice
is appropriate for comparison with star counts maps at high angular
resolution, such as HST observations, because giants are typically
masked out of the analysis \citep[e.g. see][]{dem07}. Focusing on main
sequence stars allows us to significantly reduce the surface
brightness profile fluctuations that arise from the small number of
luminous giants that would otherwise dominate the light profile.

We then select a random direction and project the main-sequence stars
positions creating a 2-dimensional map. Within this map we search for
the center of the system using the \citet{cas85} density-center method.

A circular surface brightness profile is then constructed using
$Int[\sqrt{N_{MS}}]$ annuli (where $N_{MS}$ is the number of main
sequence stars) each containing $\sim \sqrt{N_{MS}}$ sources. From
this profile we define the half-light radius in projection $r_{hP}$ as
the radius containing half of the total luminosity. The King
concentration parameter $c=\log_{10}{(r_t/r_c)}$ is determined by
fitting the surface brightness profile with a single-mass
\citet{king66} model. We use a precomputed King profile table with
uniform spacing in $W_0$ ($\Delta W_0 = 0.1$). The fit is carried out
imposing that the total luminosity and half-light radius of the
fitting model match those of the simulation snapshot within a $\leq
1\%$ deviation. From $W_0$ and $r_{hP}$ we derive $r_t$ and $r_c$. As an
alternative definition for the observed core radius we consider the
radius at which the surface brightness profile reaches half of its
central value and we indicate this quantity with $r_{\mu}$.

To construct the global mass function of the system, we proceed to
identify close pairs of stars whose light would be blended together,
similarly to the procedure described in \citet{gill08} and
\citet{pas09}. Our main aim is to treat binaries in the simulations
consistently with observations. We capture also line-of-sight
superpositions and therefore crowding. For reference we set an
HST-like resolution of 0.05 arcsec and we consider the system, of
assumed half-light radius 3 pc, at a distance of 5 Kpc from the Sun.
This implies that we consider blended those stars separated by less
than 250 AU, which includes all our primordial hard binaries. In
addition, we get a small number of apparent binaries, typically up to a
few tens per snapshot depending on the central concentration of the
system. Given a set of blended stars, we only consider the brightest
member of the ensemble as observable for the purpose of
re-constructing the stellar mass function. As the luminosity of a main
sequence star scales approximately with $M^{7/2}$, this is an adequate
approximation. With this procedure we define a catalog of observable
main sequence stars where we include stars down to
$0.2~\mathrm{M_{\sun}}$ (for reference the observed mass function of
NCG 2298 has greater than 50\% completeness at the center at this mass
limit; see \citealt{pas09}).

The stellar mass function in main sequence stars is modeled as a power
law. Its index ${\alpha}$ is obtained by maximizing the likelihood
$L$:
%%%%%%%%
\begin{equation} 
\log {L({\alpha})} = -N_{MSR} \log{\left (\frac{m_u^{{\alpha}+1}-m_d^{{\alpha}+1}}{{\alpha}+1} \right )}+{\alpha} \sum_{i=1,N_{MSR}} \log(m_i),
\end{equation} 
%%%%%%%%
where $N_{MSR}$ is the number of resolved main-sequence stars and $m_d$ and
$m_U$ are their minimum and maximum mass (here 0.2 and 0.8 M$_{\sun}$).

Based on the catalog of resolved main-sequence stars used above to
retrieve $\alpha$, we define the total observed main-sequence mass of the system $M$ as:
%%%%%%%%%%
\begin{equation}
M = \sum_{i=1,N_{MSR}} m_i.
\end{equation}
%%%%%%%%%

\subsection{3-Dimensional, Mass-Based Structural parameters}

In a numerical simulation code such as NBODY6, the tidal radius
$\sd{r_t}$ is defined based on Eq.~\ref{eq:rt}. We define as $\sd{M}
\equiv M_T$ is the total mass of the cluster (including dark remnants
such as black holes, neutron stars and white dwarfs), that is the mass
of all the particles within $2 \sd{r_t}$ from the center of the
cluster.

Several different definitions have been used in numerical simulations
for the core radius, so we direct the reader to \citet{tre07a} for
a comprehensive discussion. Here we adopt the standard definition from
\citet{cas85}: 
%%%%
\be \label{eq:rcCH}
\sd{r_c}= \frac{\sum_{i=1,N_{tot}}{m_i r_i \rho_i}}{\sum_{i=1,N_{tot}}{m_i \rho_i}},
\ee
%%%%
where $r_i$ is the distance from the center of the system and the sum
is made over all the $N_{tot}$ particles of the system (including dark
remnants). The density $\rho_i$ around each particle is computed from
the distance to the fifth nearest neighbor \citep{cas85}. In passing
we note that this is \emph{not} the NBODY6 definition of $\sd{r_c}$,
which is based on a $\rho^2_i$ weight (see
\citealt{tre07a}).
The half-mass radius of a simulated system ($\sd{r_h}$) is defined as
the tri-dimensional radius that contains half of the total mass of the
system ($\sd{M}$), thus including again any compact remnant that does not
contribute the the surface brightness profile of the system.

\section{Results}\label{sec:results}

\subsection{Mass Function Evolution}

The evolution of the mass function index $\alpha$ is shown in
Fig.~\ref{fig:alpha_m} as a function of the observed remaining mass of
the system. As time progresses (and thus the system loses stars and
mass), the mass function is preferentially depleted of low mass stars
and becomes flatter (larger $\alpha$). Models in a stronger external
tidal field lose star faster \citep{tre07a} and consequently have
larger variation in $\alpha$. However when looked at in terms of the
remaining fractional mass, the evolution of $\alpha$ becomes
independent of the tidal field strength and almost independent of the
binary fraction considered in the runs. A larger binary fraction tends
to moderately slow down the flattening of the mass function because
low mass stars can be retained, and observed, if they have a dark
remnant as a companion. All the simulations in Fig.~\ref{fig:alpha_m}
start with the same IMF below the turn-off (but with different
retention fractions of neutron stars and black holes in the right
panel). In addition, their initial conditions are such that the tidal
field cut-off is self-consistent with the initial density profile of
the system. To investigate if a universal evolution of the mass
function index is present in a broader range of conditions, we
consider in Fig.~\ref{fig:delta_alpha} additional runs that either
start from a different IMF (Salpeter) or that are initially
underfilling their Roche lobe. For the latter, we start from a $W_0=3$
King model but we set $\sd{r_t}=6.28$, which is twice as large as the
self-consistent $\sd{r_t}(W_0=3)=3.14$. To compare the evolution of
different mass functions, we consider the change in the power-law fit
of the mass function: $\Delta \alpha = \alpha(t)-\alpha(t=0)$. From
Fig.~\ref{fig:delta_alpha} it is immediately clear that the two
simulations starting with a Salpeter IMF behave very similarly to the
Miller\&Scalo runs. The run initially starting with an underfilled
Roche lobe does instead evolve marginally faster. $\Delta \alpha$ is
in fact larger at a given fraction of mass loss, especially in the
early stages of the evolution of the system. At later times the Roche
lobe is eventually filled, consistent with the results of the
simulations by \citet{gieles08}. At that point the evolution of
$\Delta \alpha$ is similar to our standard models. The close
correlation between the slope of the stellar mass function and the
fraction of the initial mass lost by a cluster was also found in
previous studies \citep{vh97,bm03}. Our simulations confirm this
relation and show its validity for a broader range of different
initial conditions. Recently, \citet{kru09} investigated the evolution
of the mass function with an analytic model, that reproduces the
universality of $\Delta \alpha$ for different IMF. The \citet{kru09}
conclusion that the retention fraction of black holes influences the
evolution of the mass function (see Fig.~15 and Sec. 5.2 in that
paper) is however not seen in our simulations. This is likely because
the analytic model of \citet{kru09} does not take into account
multiple encounters (three body or higher) of stellar black holes that
contribute to their mutual ejection on a relaxation timescale
\citep{merritt04}. The models of \citet{kru09} with different
retention fractions start to show differences in the evolution of
$\alpha$ when more than half of the mass of the system is lost. At
that point only a few stellar mass BH would have survived in the
system, independently of their initial number (e.g. see
\citealt{merritt04,gill08}).

The quasi-universal evolution of $\alpha$ as a function of $M/M(t=0)$
is very interesting because, if the initial mass function of Galactic
globular clusters is universal, as for example suggested by
\citet{Kroupa}, then the slope of mass function is a direct tracer of
the mass loss of the system, and thus of its dynamical age and of the
tidal field it experienced during its life. This diagnostic is most
effective to detect a large mass loss as $|d \alpha/d M|$ steepens
when $M$ decreases. Our conclusions on the quasi-universality of the
$\alpha$ evolution can be applied to old globular clusters, with a
turn-off mass below $1M_{\sun}$. For younger systems both the turn-off
mass and the tidal disruption timescale may introduce deviations from
the universal behavior observed here \citep{kru09}. Finally it is
interesting to note that our conclusions are consistent with Galactic
globular clusters observations of $\alpha$ \citep{dem07}: the clusters
more massive than $\sim 2\times 10^5 M_{\sun}$ have essentially the
same $\alpha$, while lower mass clusters usually show a broader range
of $\alpha$ values, likely indicating that mass loss has taken place.

\subsection{Structural parameters from King model
  fits}\label{sec:rcrh}

The evolution of the observed structural parameters also tends to
attain a universal configuration, as shown in
Figs.~\ref{fig:rcrh_m_64} and \ref{fig:rcrh_m_32} for the
core-to-half-light radius ratio and in Fig.~\ref{fig:c_m_64} for the
concentration $c$. The left panel of Fig.~\ref{fig:rcrh_m_64} shows
the $r_c/r_{hP}$ evolution of $N=64$k models with different tidal field
configurations and initial density profiles as well as a binary
fraction from 0\% to 5\%. Eventually all these models reach a common
value for the \emph{observed} $r_c/r_{hP}$ ratio as derived from King
profile fitting. 
%The long term value of this ratio slowly increases
%with time (that is as $M/M(0)$ decreases), in line with theoretical
%expectation from both numerical simulations \citep{heg06,tre07a} and
%analytical modeling \citep{ves94} that predict a larger
%$\sd{r_c/r_{h}}$ for systems with a lower number of particles. 
However the observed $r_c/r_{hP}$ ratio behaves differently from the
theoretical and numerical expectation for $\sd{r_c/r_{h}}$. In fact,
core-collapse for systems made only of single stars is detectable only
for a very short time based on $r_c/r_{hP}$. After core-collapse, runs
with 10\% primordial binaries, (left panel of
Fig.~\ref{fig:rcrh_m_32}) behave essentially as those with single
stars only. We interpret this result as a consequence of mass
segregation of dark remnants at the center of the cluster. The
dark-remnants are on average heavier than the main sequence stars and
thus they transfer kinetic energy to them. This results in a heating of
the visible component of the cluster, which therefore does not exhibit
a typical core-collapsed structure. An independent confirmation of
this trend can be inferred from Fig.~5 in \citet{hurley07}, where runs
with and without primordial binaries also have a similar core
(although \citealt{hurley07} runs do not reach past core-collapse).
The core collapse for single-star systems is instead well discernible
from the evolution of $\sd{r_c/r_{h}}$ (see Fig.~\ref{fig:rcrh_comp}).
Until core-collapse, theoretical and observed King-fitted core radii
evolve similarly, but immediately after the core bounce the two
definitions separate sharply from each other. Post core-collapsed
systems develop a shallow central cusp ($\mu \sim R^{-\nu}$ with $\nu
\sim 0.4-0.7$) that is missed by the best fitting King model (see
Fig.~\ref{fig:sb_fit}), whose structure is constrained by the global
surface brightness profile. In general, a King model becomes a poorer
description of the system after core-collapse: in fact, the quality of
our fits is degraded by about a factor two, as measured by the ratio
of typical $\chi^2$ before and after core-collapse. As a consequence,
the fit results in the post core-collapsed phase also become somewhat
dependent on the modeling of photometric errors. We assumed a constant
fractional error on the surface brightness profile, which corresponds
to a constant error in magnitude, except for the outer points, where
we assigned a constant error in flux (to model the effect of the sky
background). Assuming a constant error in flux for all the points in
the surface brightness profile yields to no differences in the fit
before core collapse. After the collapse, slightly more concentrated
models are preferred, because in this case the points with the highest
surface brightness have an increased weight relative to the outer points.

The results in Fig.~\ref{fig:rcrh_m_64} have been obtained assuming a
\citet{ms} IMF and a 100\% retention of neutron stars and stellar-mass
black holes. A \citet{salp} IMF has a higher fraction of dark remnants
per unit mass (because of it shallower slope, $\alpha = -2.3$,
compared to the MS IMF, $\alpha=-3.3$, in the range $[10:100]
M_{\sun}$), thus it sustains a larger core, especially in the earlier
stages of the simulation (see right panel of
Fig.~\ref{fig:rcrh_m_32}). This effect is primarily due to stellar
mass black-holes \citep{merritt04,mackey08}, so $r_c/r_{hP}$ returns in
line with that of the \citet{ms} simulations once the BHs segregate at
the center of the cluster and kick each-other out of the system via
three body interactions. Reducing the retention fraction of neutron
stars and BHs to 30\% leads to a decrease of the core radius down to
$r_c/r_{h} \sim 0.05$, albeit with large fluctuations. The large
fluctuations that appear after the core-collapse are also related to
the degraded fit-quality, which increases the uncertainty on $c$ (that
is on $W_0$).

Interestingly, the main-sequence surface brightness profile has a
significantly larger core in the run with a central IMBH compared to
all other runs once they reach their long-term equilibrium
configuration (see right panel of Fig.~\ref{fig:rcrh_m_32}). The
prediction of \citet{tre07b} and \citet{gill08} that collisionally
relaxed systems with a central IMBH should exhibit a large $r_c/r_{hP}$
is thus confirmed (see also \citealt{baum05}). However systems with a
large core radius might also be in the pre-core collapse stage. Unless
the cluster relaxation time is short enough to suggest the cluster
should have already undergone core collapse, no IMBH (or IMBH-like
heating) is required to explain large values of $r_c/r_{hP}$. A small
value of $r_c/r_{hP}$, on the other hand, can be safely used to exclude
likely candidates to harbor an IMBH.

Fit of the surface brightness profile with a King model does not
appear to provide a very solid determination of the intrinsic tidal
radius of the system. The comparison between $r_t$ and {$\sd{r_t}$} in
Fig.~\ref{fig:tidal} shows that, when the entire surface brightness
profile is used for the fit, then $r_t > {\sd{r_t}}$. This is however
not surprising, because unbound particle escaping from the system are
removed from the calculation only when they have $r>2 \sd{r_t}$, thus
the surface brightness profile at radii beyond $\sd{r_t}$ may still be
positive. Similarly in actual star clusters, unbound stars remains
initially in the proximity of the system, while escaping. The measure
of $r_t$ becomes more uncertain, and dependent upon the details of the
fit, if only a fraction of the surface brightness profile points is
used (see blue dotted line in Fig.~\ref{fig:tidal}). This is
especially true after core-collapse (that is at $M/M(0) \lesssim 0.65$
in Fig.~\ref{fig:tidal}. This is primarily because we carried out the
fit over the full-surface brightness profile by fixing the total
luminosity and the half-light radius within 1\% of the values in the
profile. If only a fraction of the points is available, these two
scales cannot be measured directly and must be left as free parameters
of the fit. This leads to increased uncertainty, especially after core
collapse, when the surface brightness profiles might present some
departures from a King profile (for example see
Fig.~\ref{fig:tidal_fit} at $R\sim 0.15$). When we do not use the
outer points in the fit, the tidal radius tends to be moderately
underestimated after core collapse compared to the determination using
the full profile (Fig.~\ref{fig:tidal}). This is consistent with what
has been found in the determination of the tidal radius for some
Galactic globular clusters (\citealt{mcl05}). A different conclusion
may however be reached for those globular clusters that do not fill
their Roche lobe \citep{bau09}.

\subsection{Alternative core radius definitions}

Given the difficulty in reliably identifying core-collapsed systems
from King model fitting, we explore here alternate definitions for the
observed core radius. One possibility is to use the classical core
radius $r_{\mu}$, that is the radius at which the surface brightness
falls to half of its central value. A comparison between $r_c$ and
$r_{\mu}$ is shown in Fig.~\ref{fig:rc_defs} and suggests that this
definition is better suited for identifying core-collapsed systems.
$r_{\mu}$ remains in fact close to $\sd{r_c}$ (within $\sim 15\%$
accuracy), confirming for complex dynamic configurations that include
mass segregation the results published by \citet{cas85} for King
profiles with a constant mass-to-light ratio. However the definition
of $r_{\mu}$ depends on the choice of bins, especially on the central
one. Too small a bin size will lead to large Poisson fluctuations in
the central surface brightness value, while an excessively large bin
will overestimate the core size by lowering the central value of the
surface brightness. A large central bin is not necessarily a choice,
but may be imposed in actual observations by a limited angular
resolution. To overcome some limitation related to binning,
\citet{cas85} proposed to extend the density-based core radius
estimator to the surface brightness profile $\mu(R)$ and to define a
surface-brightness density radius $s_{\mu}$. In our notation this
reads:
%%%%%%%%%%
\be \label{eq:rmuCH}
s_{\mu}= \frac{\sum_{i=1,N_{MSR}} \mu(R_i) R_i }{\sum_{i=1,N_{MSR}} \mu(R_i) },
\ee
%%%%%%%%%%
where the sum is carried out over all the detected main-sequence
stars. $s_{\mu}$ is shown for our simulations in
Fig.~\ref{fig:rc_defs}. As expected from Fig.~3 in \citet{cas85},
$s_{\mu}$ deviates from $r_{\mu}$ and $\sd{r_c}$ for
high-concentration King models, that is mainly in the post
core-collapsed phase. Interestingly the value of $r_c$ obtained from
King fitting for post core-collapsed systems is approximately bound
between $r_{\mu}$ and $s_{\mu}$.

\subsection{Fluctuations}

Fluctuations in the structural parameters beyond those expected from
Poisson noise are clearly apparent from our analysis. \citet{heggie09}
reach the same conclusion for their simulation of NGC 6397. Our run
with a \citet{ms} mass function and low retention of dark remnants has
particularly strong fluctuations, as can be seen for all different
core radius estimators in Fig.~\ref{fig:rc_defs}. It is very
interesting to note that the system appears to go through some
breathing phases, where subtle changes are made to the surface
brightness profiles, affecting primarily the King model fit. The
origin of these fluctuations is likely dynamical, especially
because we are considering only the main-sequence stars and thus we
are relatively unaffected by low-number statistics. 

\subsection{Identification of core-collapsed systems}

Our results on the determination of core radii have profound
implications for our understanding of the dynamical state of observed
globular clusters: most intrinsically core-collapsed globular clusters
in our simulations appears to have a surface brightness profile in
main sequence stars essentially identical to non-core collapsed
systems when a King model is used to retrieve $r_c$. Furthermore our
simulations do not predict very small $r_c$ if the retention fraction
of neutron stars and black holes is high. Alternative definitions of
the core radius, such as $r_{\mu}$, are more efficient at identifying
core-collapsed clusters, but they may still suffer from large
fluctuations and/or biases due to finite resolution at the center of
the system. This picture clearly illustrates the potential problems in
the observational identification of core collapsed clusters. In fact,
the classification of an observed globular as core-collapsed or not
might be a result of the combination of significant fluctuations in
the total surface brightness profiles (enhanced with respect to our
main-sequence only analysis because the profile is defined primarily
by a small number of giant stars; see also \citealt{heggie09}),
combined possibly with the limited angular resolution of ground-based
data \citep{trager95} used to construct the structural parameters
presented in the \citet{har96} catalog. The core size in the post
core-collapse phase depends mildly ($\propto 1/log(0.11 N)$) on the
number of particles in the system \citep{ves94,heg06}, thus it is expected
that smaller core radii can be measured in systems with a larger
number of particles. Further exploration of this issue is deferred to
a follow-up paper that will include live stellar evolution to follow
stars off the main sequence and thus to correctly model the total
luminosity of the cluster. This will allow us to carry out a more
direct comparison with the ground-based data that define $r_c$ in the
\citet{har96} catalog.

\subsection{Evolution in the $(c;\alpha)$ plane}

An interesting correlation has been recently reported by \citet{dem07}
between the slope $\alpha$ of the global stellar mass function of
globular clusters and their central concentration $c$. Less
concentrated clusters are more depleted in low mass stars than those
with higher values of $c$.
%\footnote{This trend is opposite to what has
 % been suggested in \citet{mcl08}.}

The evolution of a star cluster in the $(c;\alpha)$ plane is driven by
collisional (two-body) processes, thus the \citet{dem07} observations
test our understanding of the dynamics of globular clusters. As
discussed in Sec.~\ref{sec:intro}, the observed correlation appears to
be in conflict with the standard scenario globular cluster evolution
which predicts that relaxation drives an increase in $c$ as the
cluster evolves toward core collapse while at the same time
preferentially depleting the system of low mass stars via evaporation
\citep{spi87}. The evolution of our simulated star clusters in the
concentration-slope of the mass function ($c-\alpha$) plane is shown
in Fig.~\ref{fig:c_alpha}, where we also report the values observed by
\citet{dem07} for those clusters that are well relaxed ($t_{rh} \leq
1$ Gyr) according to the \citet{har96} catalog. In plotting the
concentration parameter $c$ for the clusters shown in
Fig.~\ref{fig:c_alpha} we have used the values obtained by
\citet{mcl05}, which are based on the \citet{king66} models.
Differences from the concentration values reported in \citet{dem07}
are however very small ($\Delta c < 0.1$, except for NGC 6712 which
has $c=1.05$ in \citet{mcl05} and $c=0.9$ in \citet{dem07}). While our
simulations never reach a deep core-collapse ($c \gtrsim 2.5$), their
concentration tends to evolve toward $c\sim 1.7$, still larger than
what observed for the systems with the most depleted stellar mass
function. One possibility to explain the system with lower
concentrations remains that of a very efficient source of heating at
the center of the system, comparable with the heating generated by an
IMBH (see cyan line in Fig.~\ref{fig:c_alpha}). An IMBH itself appears
to be excluded as the preferred explanation, because NGC 2298, where a
central IMBH is rejected at high confidence level \citep{pas09}, is
one of the low $c$ points in Fig.~\ref{fig:c_alpha}. Stellar mass
black-holes are also very unlikely to be viable, because the heating
source needs to remain efficient in the latest stages of the evolution
of the system. Stellar mass BHs lose instead efficiency at later times
(see red line in the lower panels of Fig.~\ref{fig:rcrh_comp}; see
also \citealt{merritt04}). It has been recently suggested that white
dwarf kicks can sustain a large core \citep{fre09}. However the effect
of the kick has been shown only at the level of $\sd{r_c/r_h}$ and
thus its full impact on the observed concentration $c$ is an open
question.

An alternative explanation for the discrepancy in the $(c;\alpha)$
plane evolution might rest in observational biases. While we carried
out our analysis restricting to main-sequence stars, the tidal radius
of observed globular clusters is mainly derived from ground-based data
\citep{trager95}, that thus are mostly dominated by giant-branch
stars. Therefore mass segregation might lead to a systematic
underestimate of $r_t$ from these observations compared to our
analysis. 

Overall Fig.~\ref{fig:c_alpha} shows that the correlation between
$\alpha$ and $c$ observed by \citet{dem07} is still in mild
disagreement with the expectations from numerical simulations. Half of
the 8 observed clusters are consistent with our models (including NGC
2298 which has $(c=1.31;\alpha=0.5)$ and could be explained by a core
fluctuation). Two of them appear instead to be core-collapsed
clusters. They have a conventional concentration $c=2.5$, higher than
our what we obtain from our simulations, as discussed in Section
\ref{sec:rcrh} (but they could be explained by assuming a low
retention fraction of neutron stars and black holes). Two clusters
have instead concentrations below our predictions, but here the
consistency of simulations and observations might improve if data and
models are analyzed with a fully self-consistent approach.

\section{Conclusion and Discussion}\label{sec:conc}

In this paper we discussed the evolution of the global mass function
index $\alpha$ and of the structural parameters for multimass direct
N-body models of star clusters in a galactic tidal field with and
without primordial binaries using a total number of particles up to
$N=65536$. We adopted a \citet{ms} initial mass function evolved with
an instantaneous step of stellar evolution to a turn-off mass of $0.8$
M$_{\sun}$. The simulation results have been interpreted with great
care in order to use definitions of structural quantities consistent
with observations of Galactic star clusters. 

\begin{figure}
 \plottwo{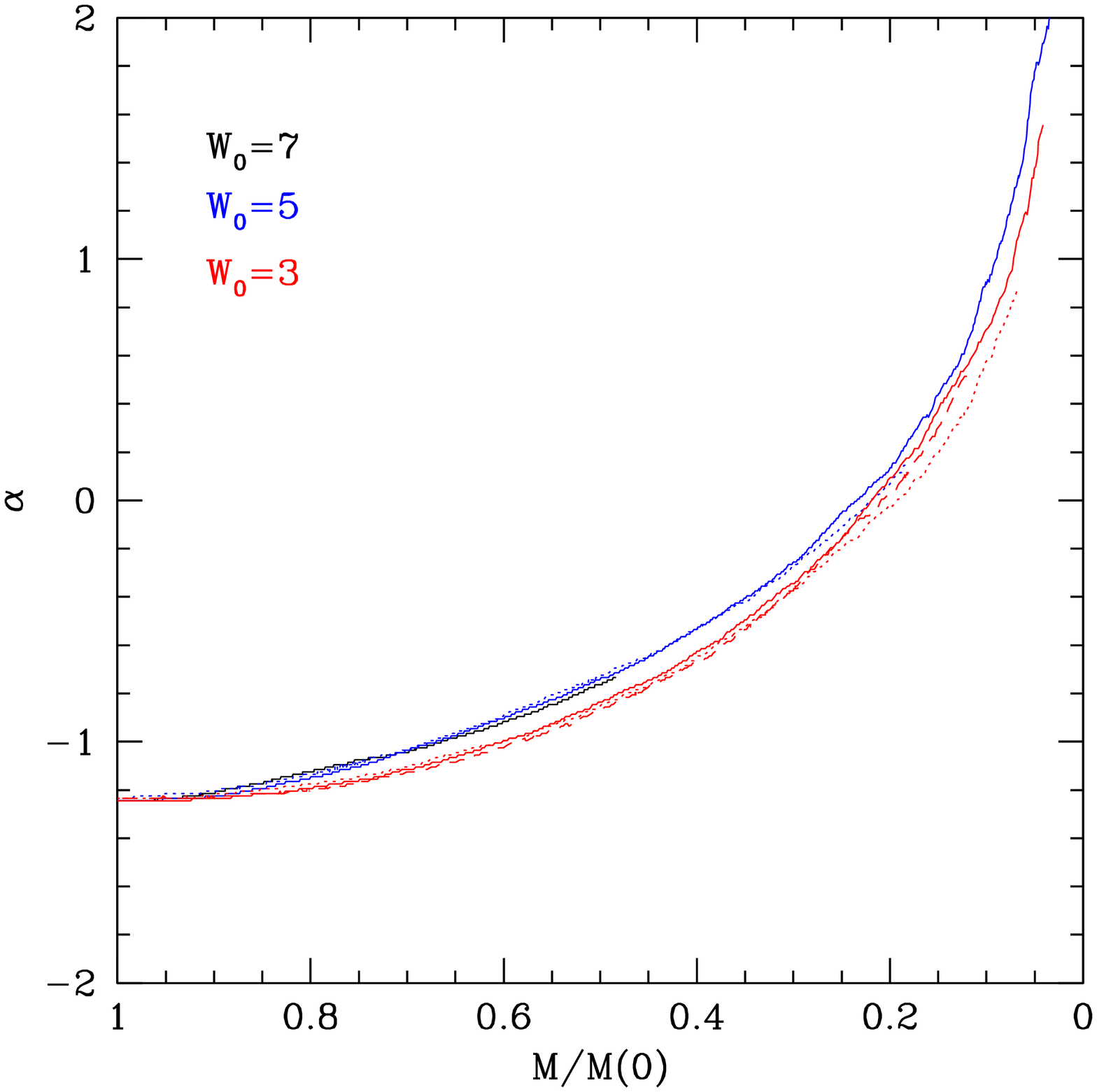}{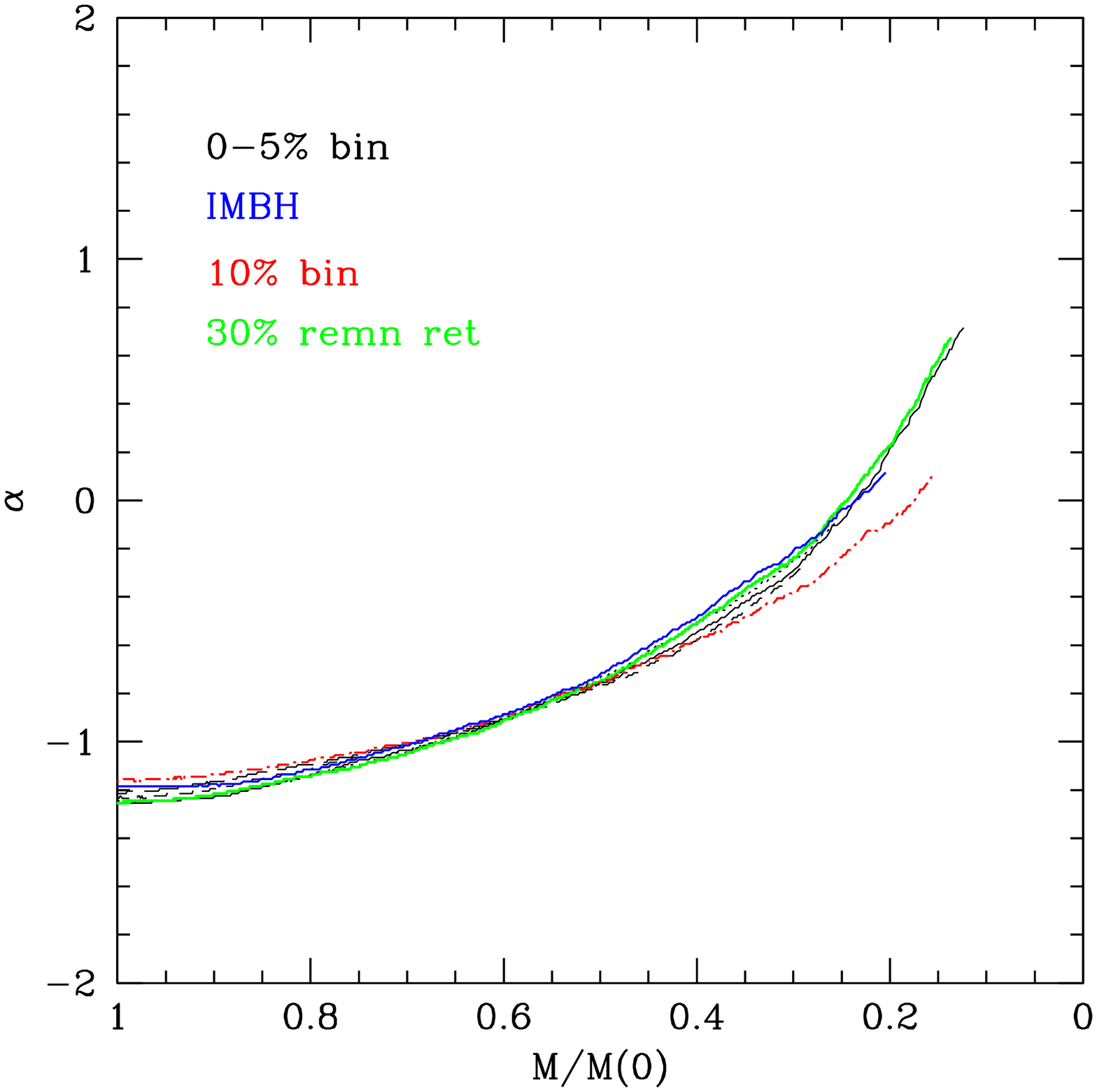}\caption{Time evolution of the
   mass function index $\alpha$ for \citet{ms} runs as a function of the total observable
   mass remaining in the system $M$. Left panel: models with
   $N_{tot}=65536$ with different tidal fields and initial density
   profiles (King models with $W_0=3,5,7$ --- red, blue, black
   colors)
   and 0,2,5 \% primordial binaries (solid, dotted, dashed lines).  Right panel: $W_0=7$ King models starting with $N_{tot}=32768$ 
   particles and a binary fraction up to 10\%. Black lines are a binary fraction 0-5\%, the red line has f=10\%. A model with no primordial binaries and a central IMBH is also shown in this panel (blue). The green line shows the model with 30\% retention fraction of neutron stars and black holes. The evolution of the mass function slope is almost universal. 
 }\label{fig:alpha_m}
\end{figure}

\begin{figure}
  \plotone{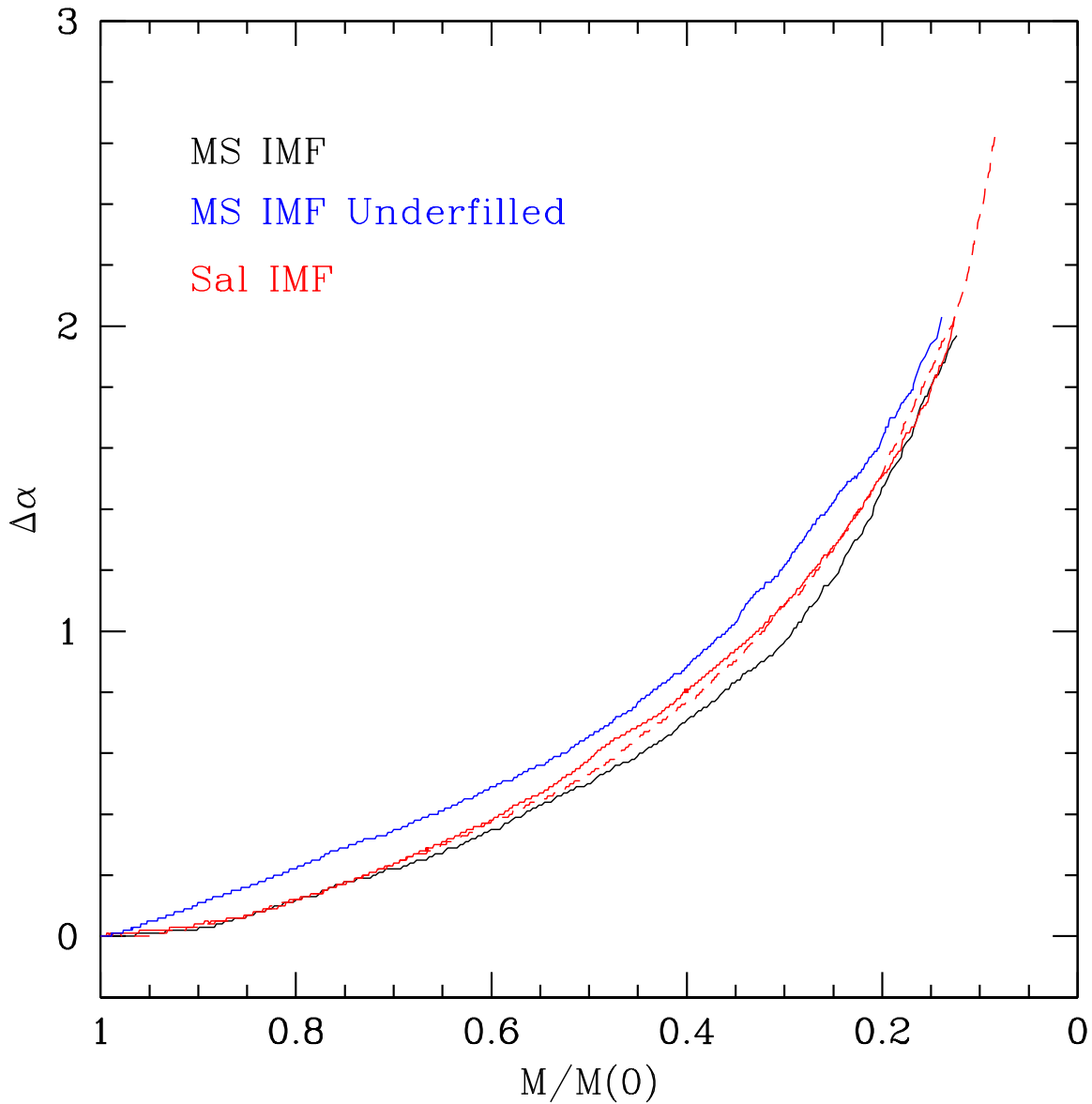}\caption{Time evolution of the mass
    function index $\alpha$, shown as the difference from its $t=0$
    value, $\Delta \alpha = \alpha(t)-\alpha(t=0)$, for a series of
    runs with a Miller \& Scalo and a Salpeter
    mass function. The black line refers to a Miller \& Scalo run with 
100\% retention fraction of dark remnants and no binaries starting from $W_0=7$. 
The blue line has the same IMF but starts a $W_0=3$ King model and $\sd{r_t}=6.28$ (this model has its Roche lobe underfilled by a factor 2).
 The red  lines are runs with a Salpeter IMF, starting from $W_0=7$ and no binaries: (solid: 100\% retention fraction,
 dashed: 30\% retention fraction). 
 All simulations have $N_{tot}=32768$.}\label{fig:delta_alpha}

\end{figure}

\begin{figure} \plottwo{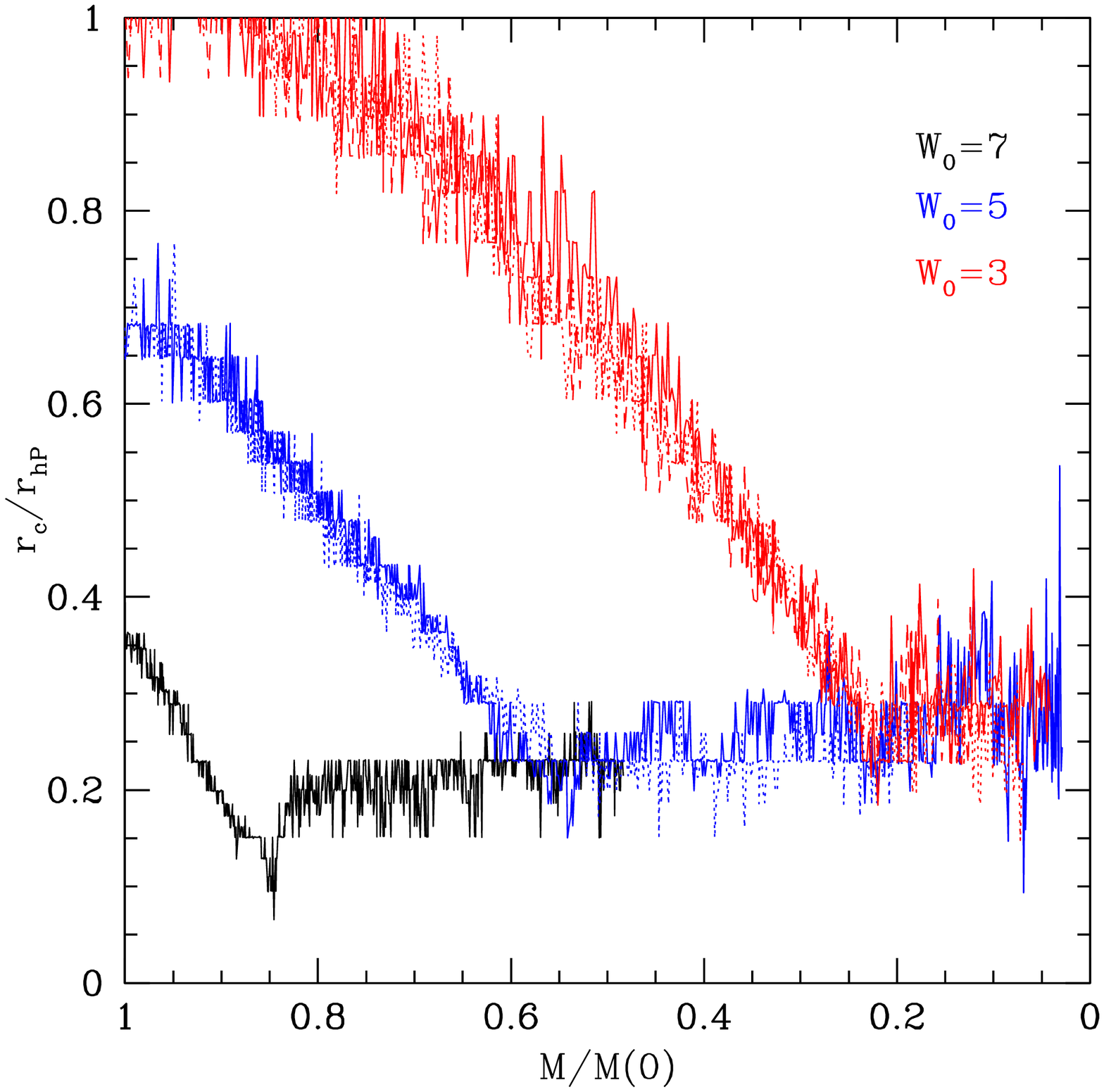}{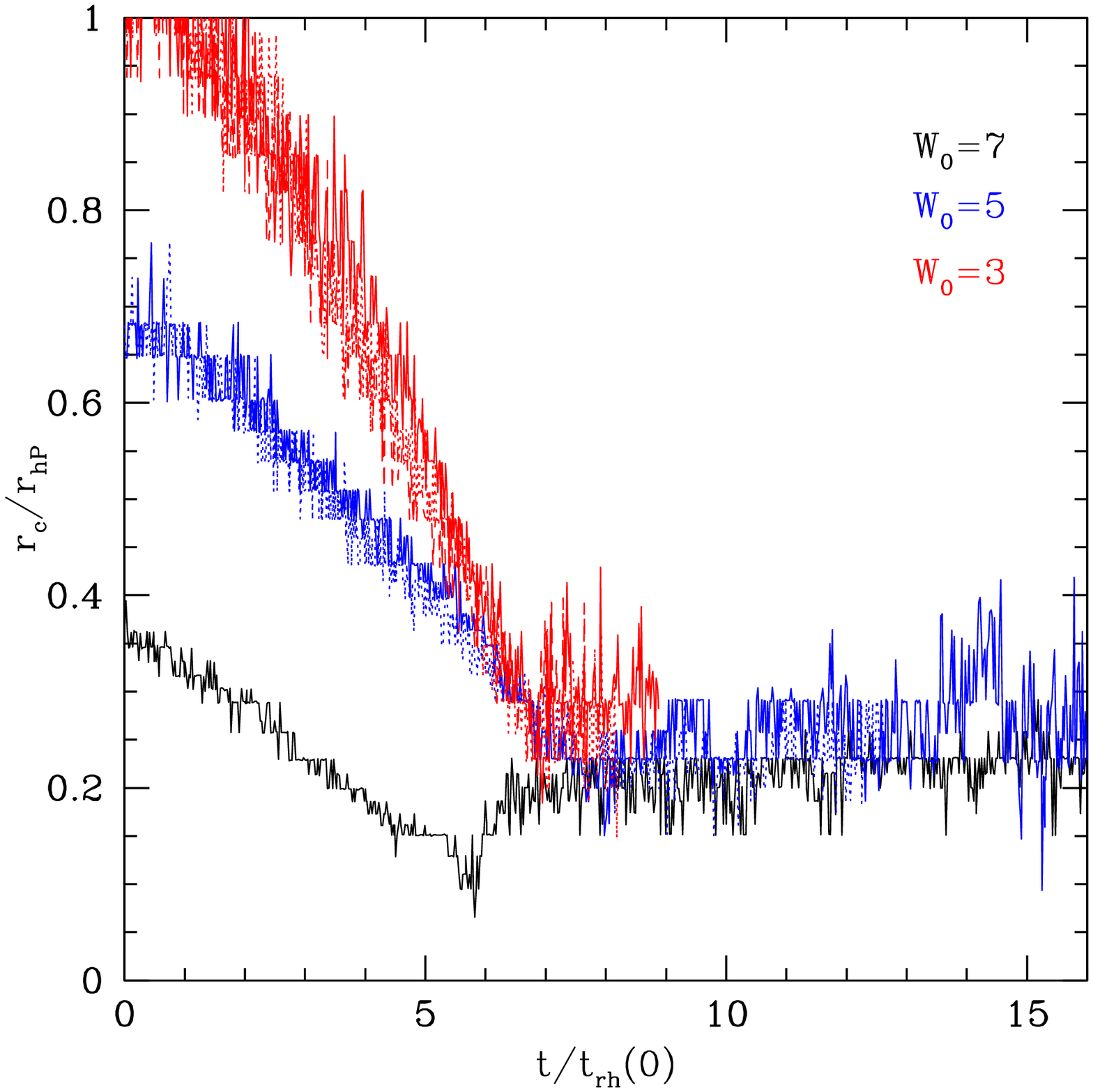}\caption{Evolution
  of the observed core to half light radius for our $N=64k$ runs, color codes
  like in the left panel of Fig.~\ref{fig:alpha_m}. Left panel shows
  $r_c/r_{hP}$ as a function of $M$, right panel as a function of $t$ in
  units of the initial half-light relaxation time.}\label{fig:rcrh_m_64}
\end{figure}

\begin{figure}
\resizebox{250pt}{!}{\includegraphics{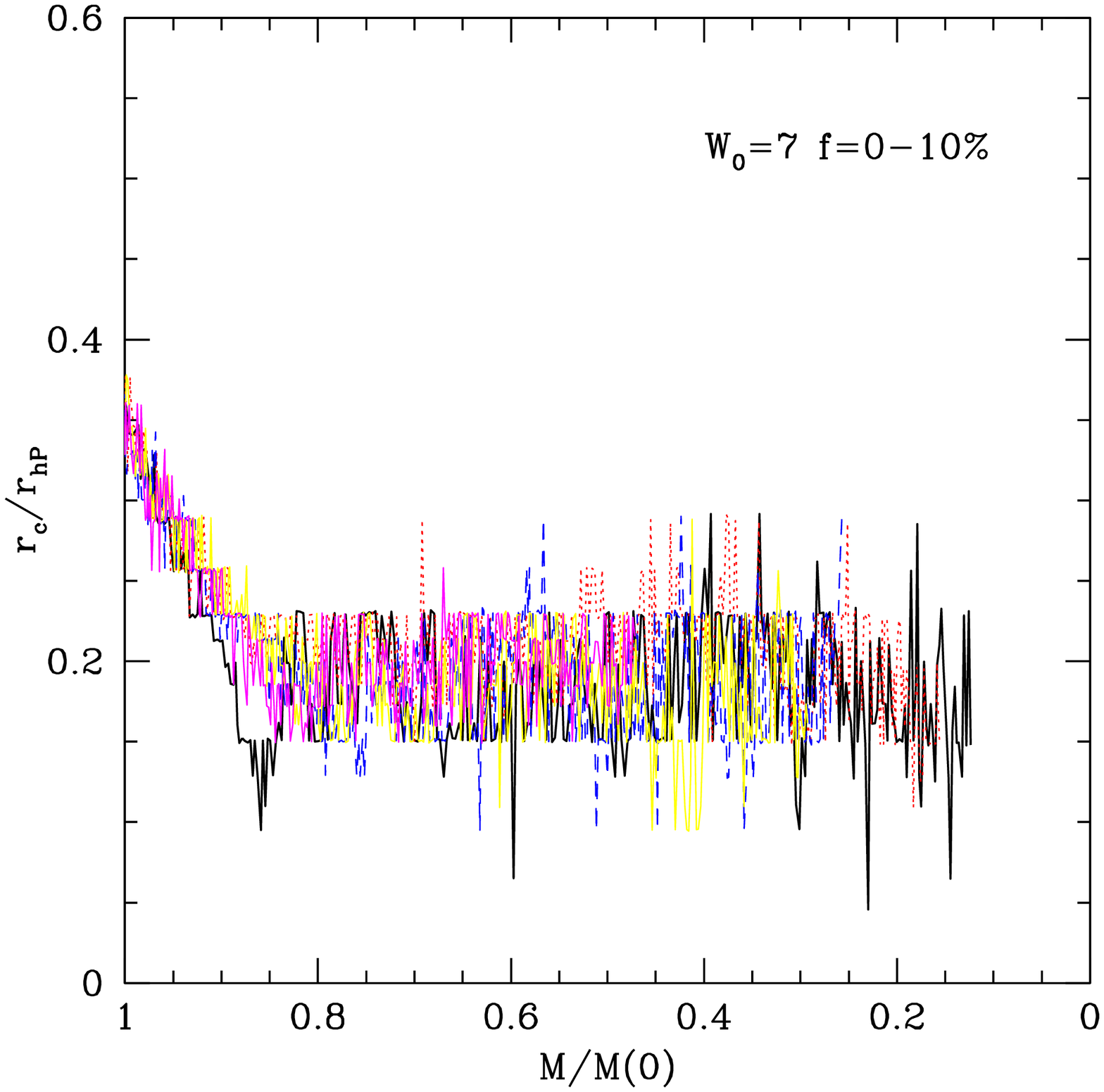}}
\resizebox{250pt}{!}{\includegraphics{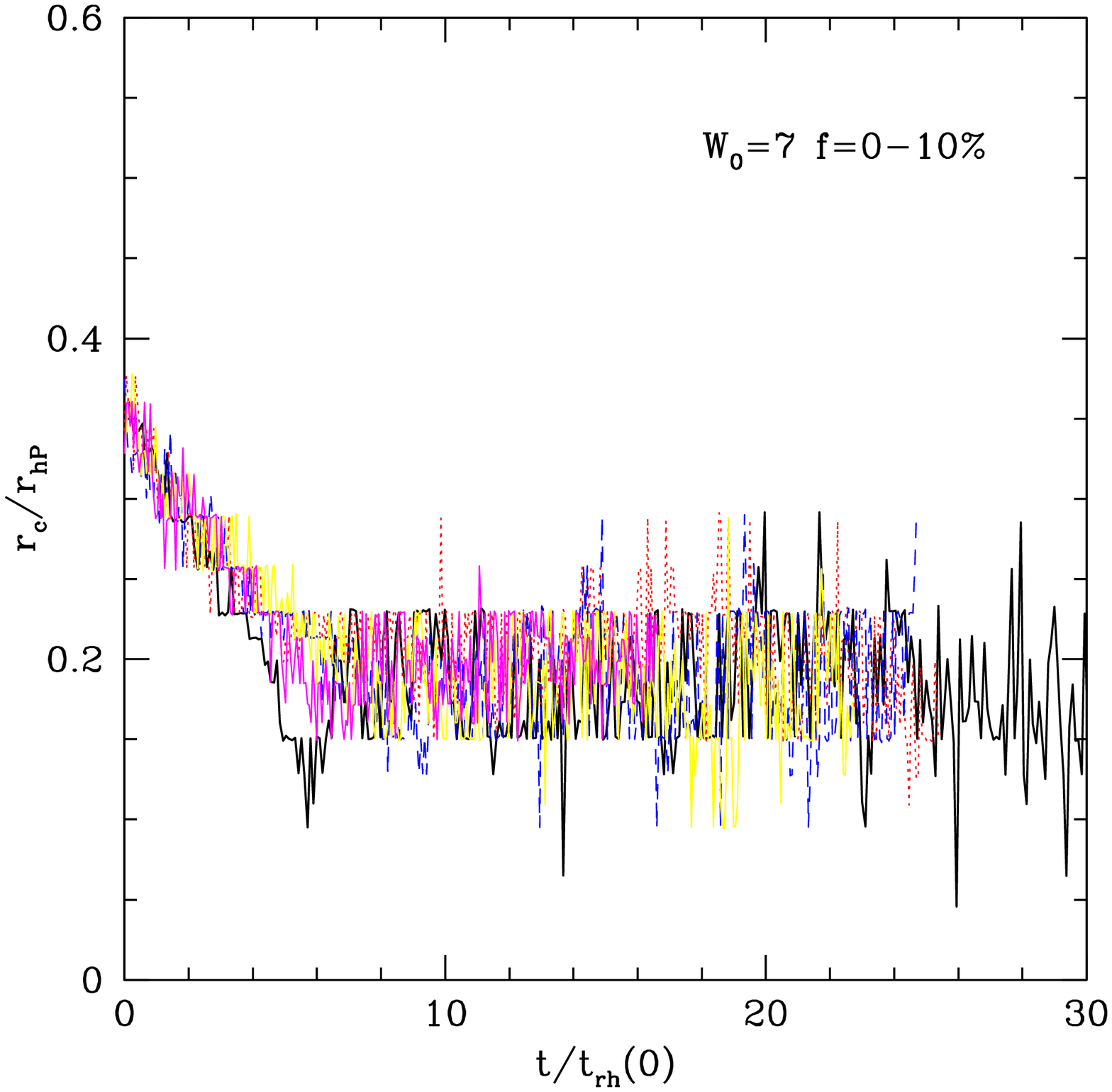}}
\resizebox{250pt}{!}{\includegraphics{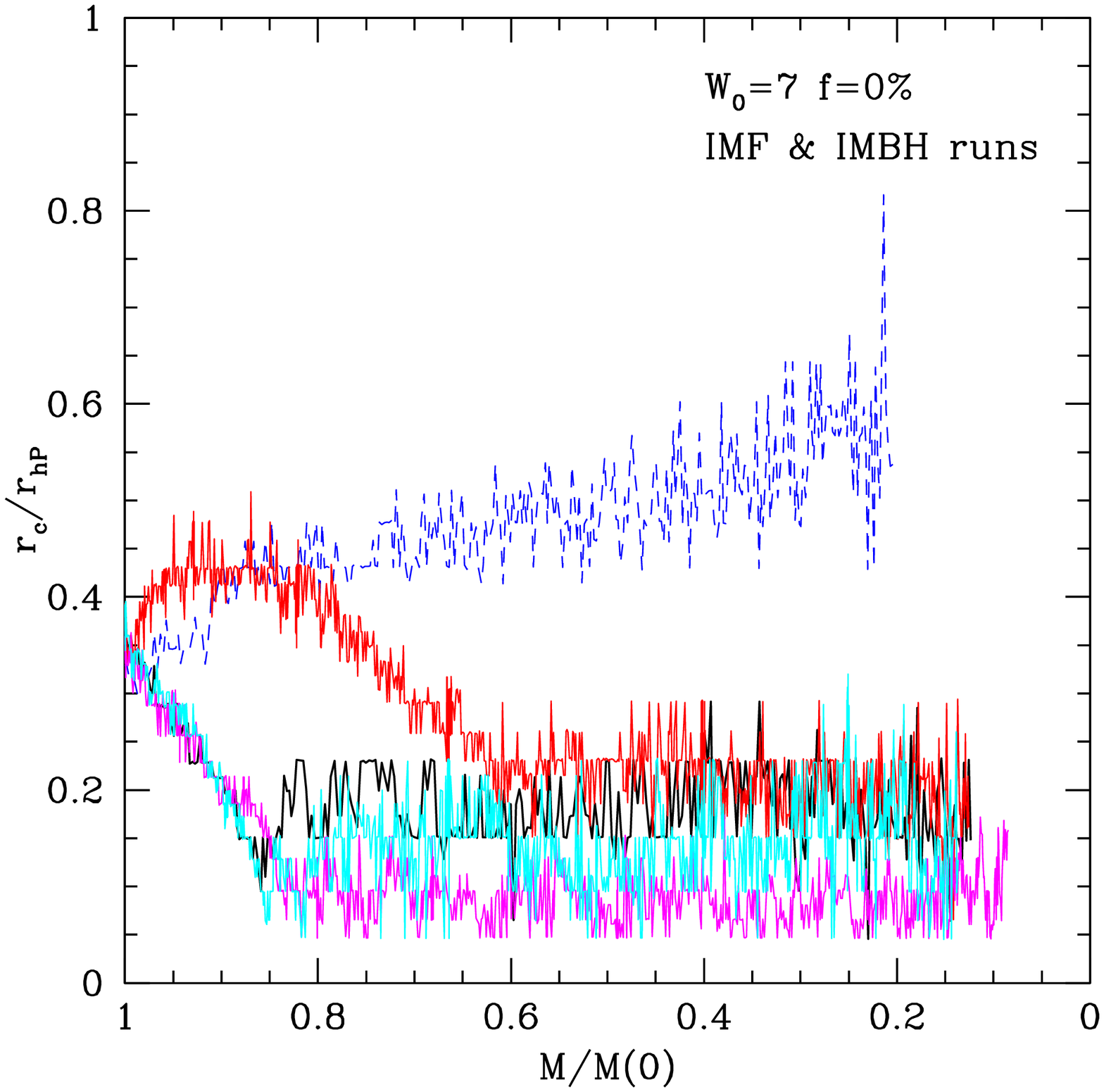}}
\resizebox{250pt}{!}{\includegraphics{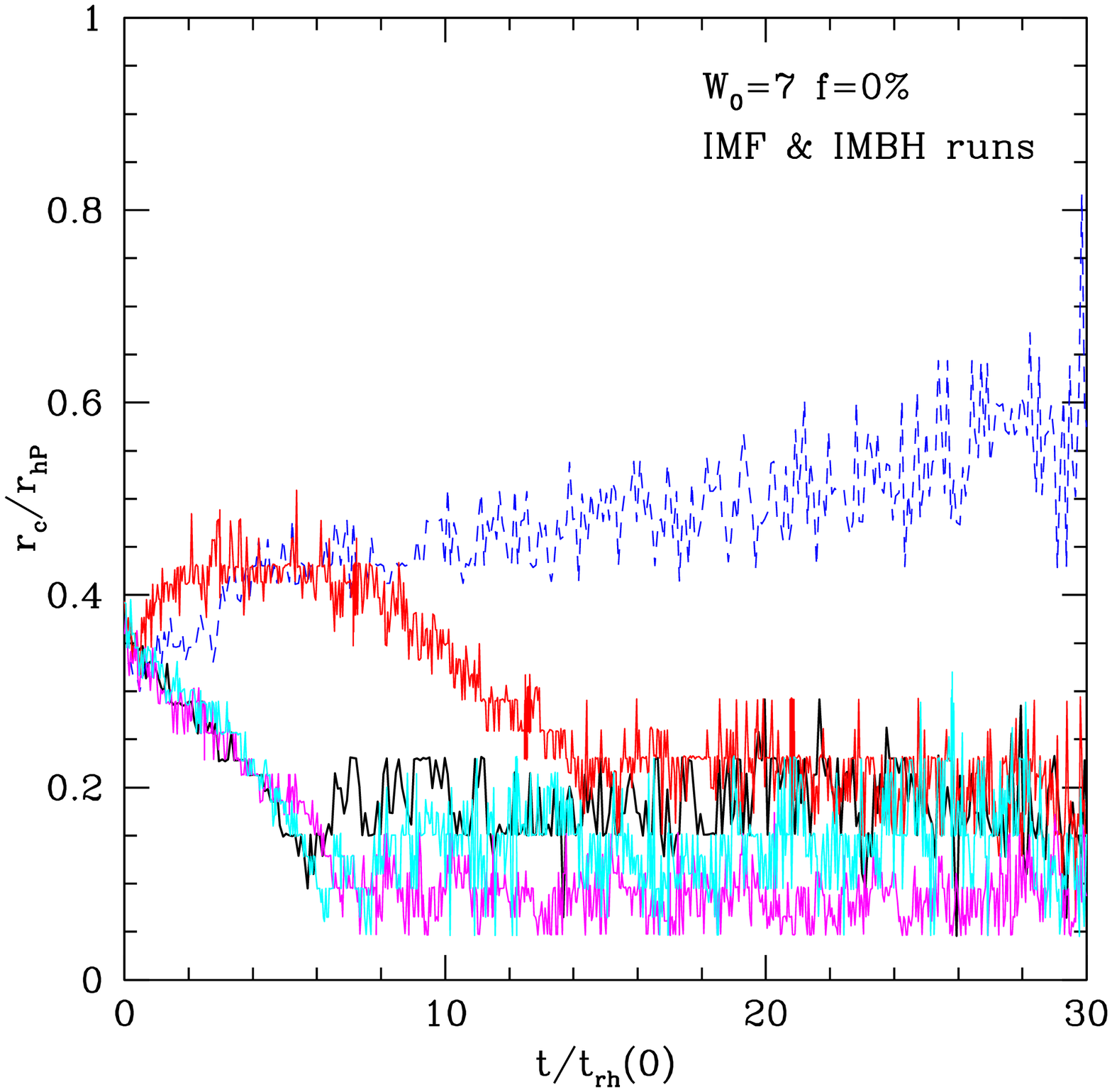}}
\caption{
    Evolution of $r_c/r_{hP}$ like in Fig.~\ref{fig:rcrh_m_64} but for
    models with $N=32k$. Upper set: \citet{ms} IMF and different
    binary fractions (0\% black, 1\% blue, 3\% yellow, 5\% magenta,
    10\% red). Lower set: models with no binaries but different IMFs. Black line: \citet{ms} with full retention of remnants, red line \citet{salp}
 with full retention of remnants. Cyan and magenta lines are the corresponding IMFs with 30\% retention fraction. Blue is our run with a central IMBH. Only
  a central IMBH is able to sustain a large $r_c$ in the
  long term, but a significant number of stellar-mass BHs can also
  give a transient high value of $r_c$. Small $r_c$ values are instead
  possible with a low retention fraction of dark
  remnants.}\label{fig:rcrh_m_32} \end{figure}

\begin{figure} \plottwo{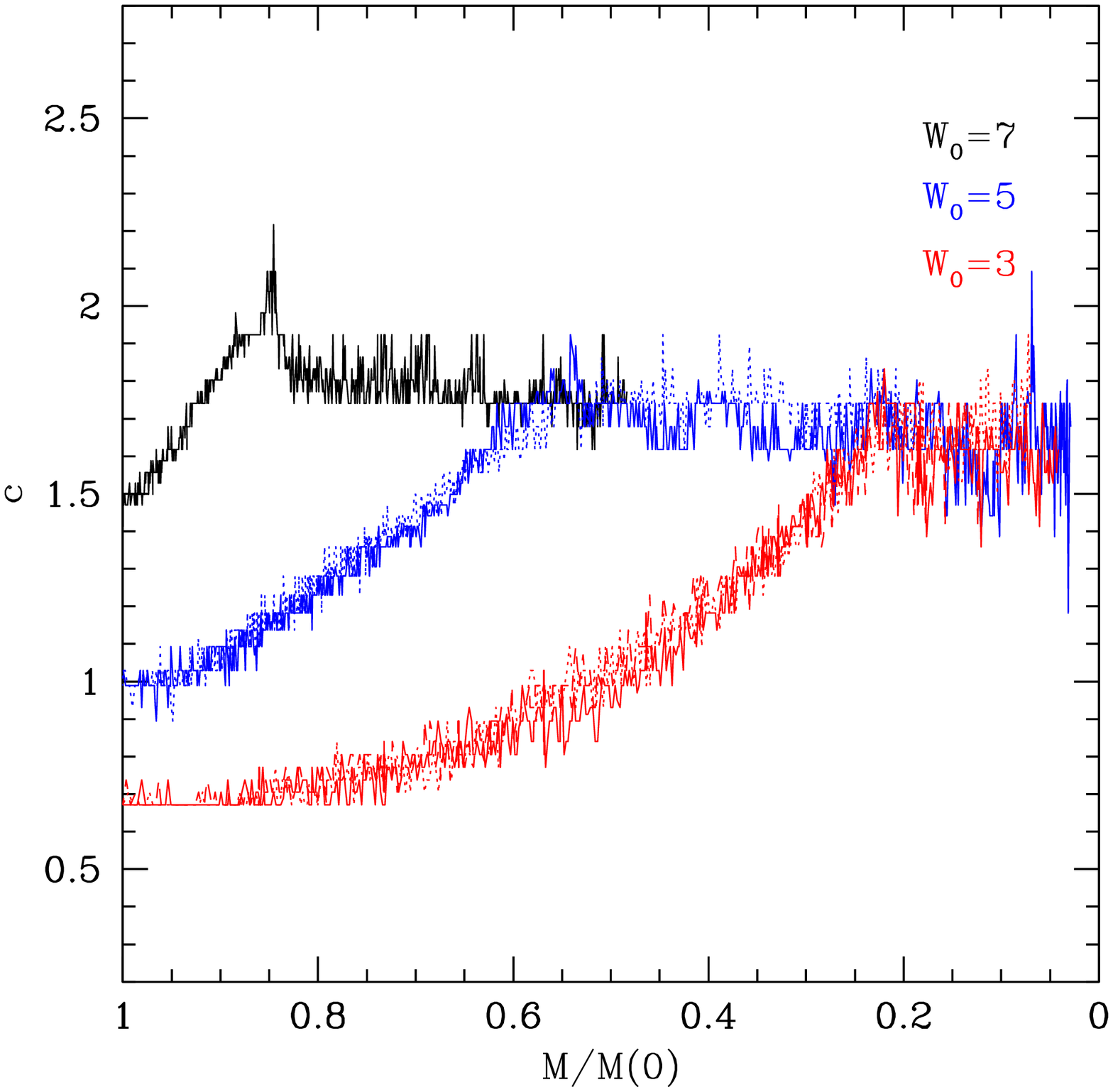}{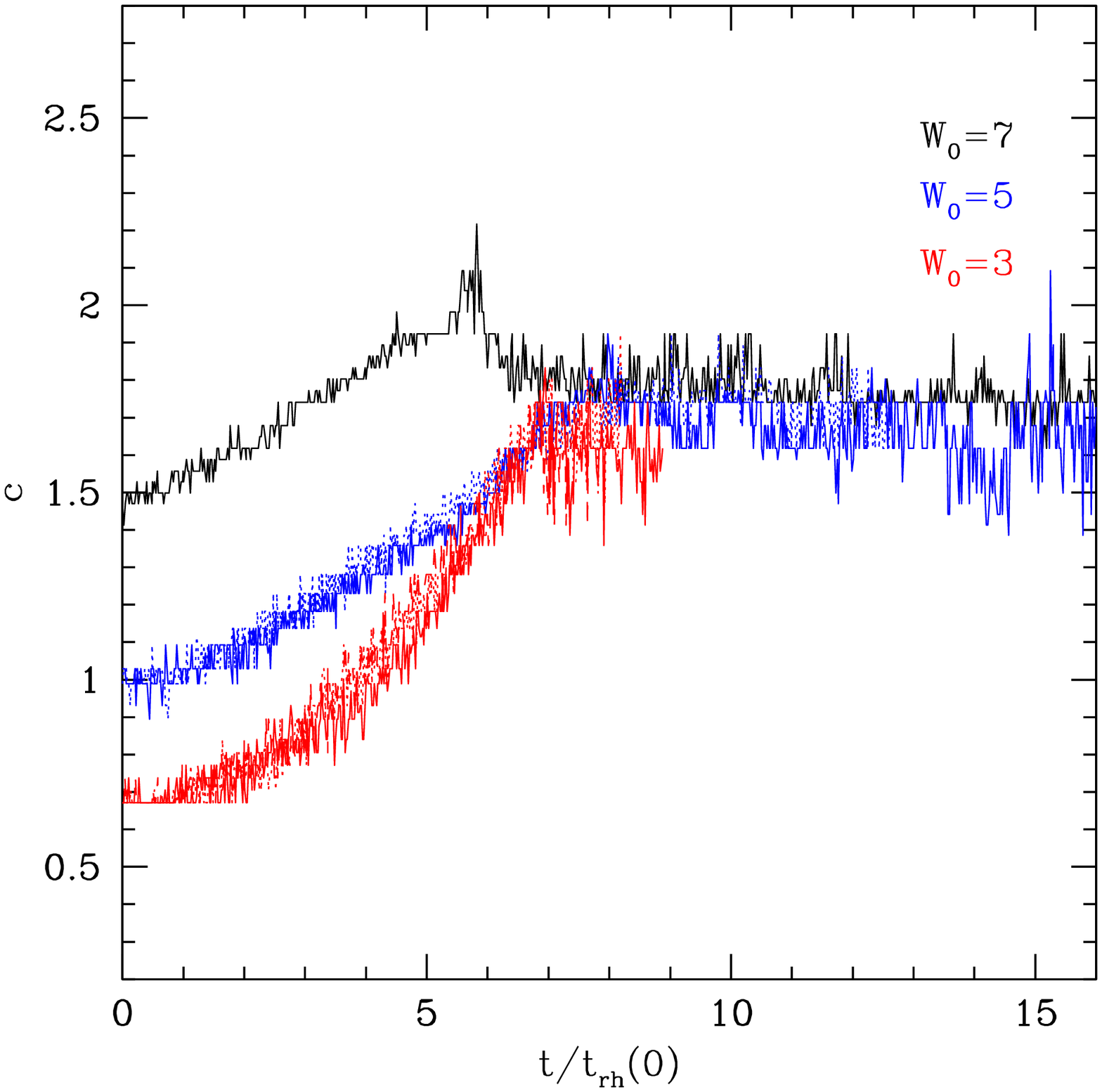}\caption{Evolution
 of the observed concentration $c$ for our $N=64k$ runs, color codes
 like in Fig.~\ref{fig:alpha_m}. Left panel shows
 $c$ as a function of $M$, right panel as a function of $t$ in
 units of the initial half-light relaxation time.}\label{fig:c_m_64}
\end{figure}

\begin{figure} 
\resizebox{250pt}{!}{\includegraphics{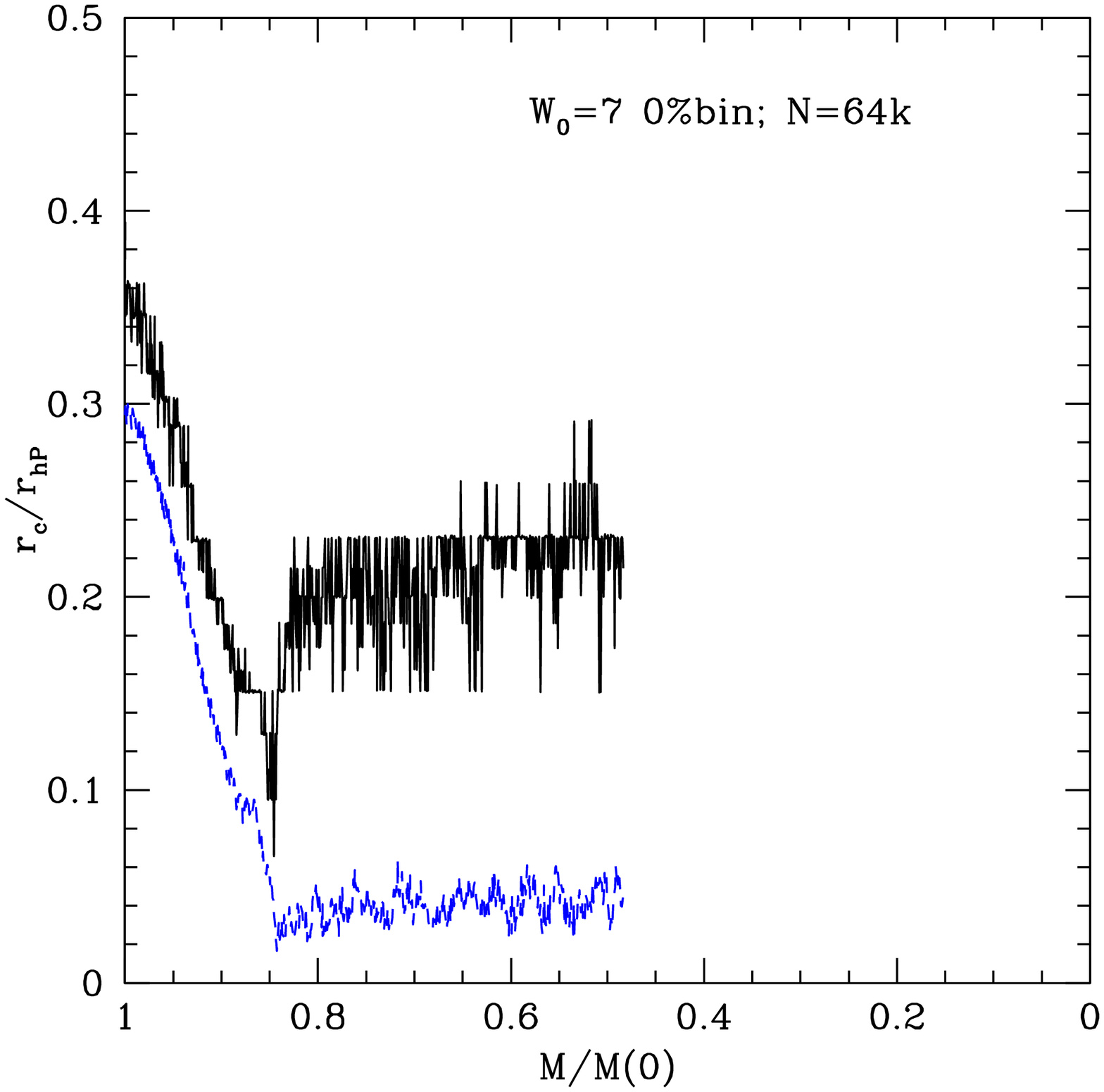}}
\resizebox{250pt}{!}{\includegraphics{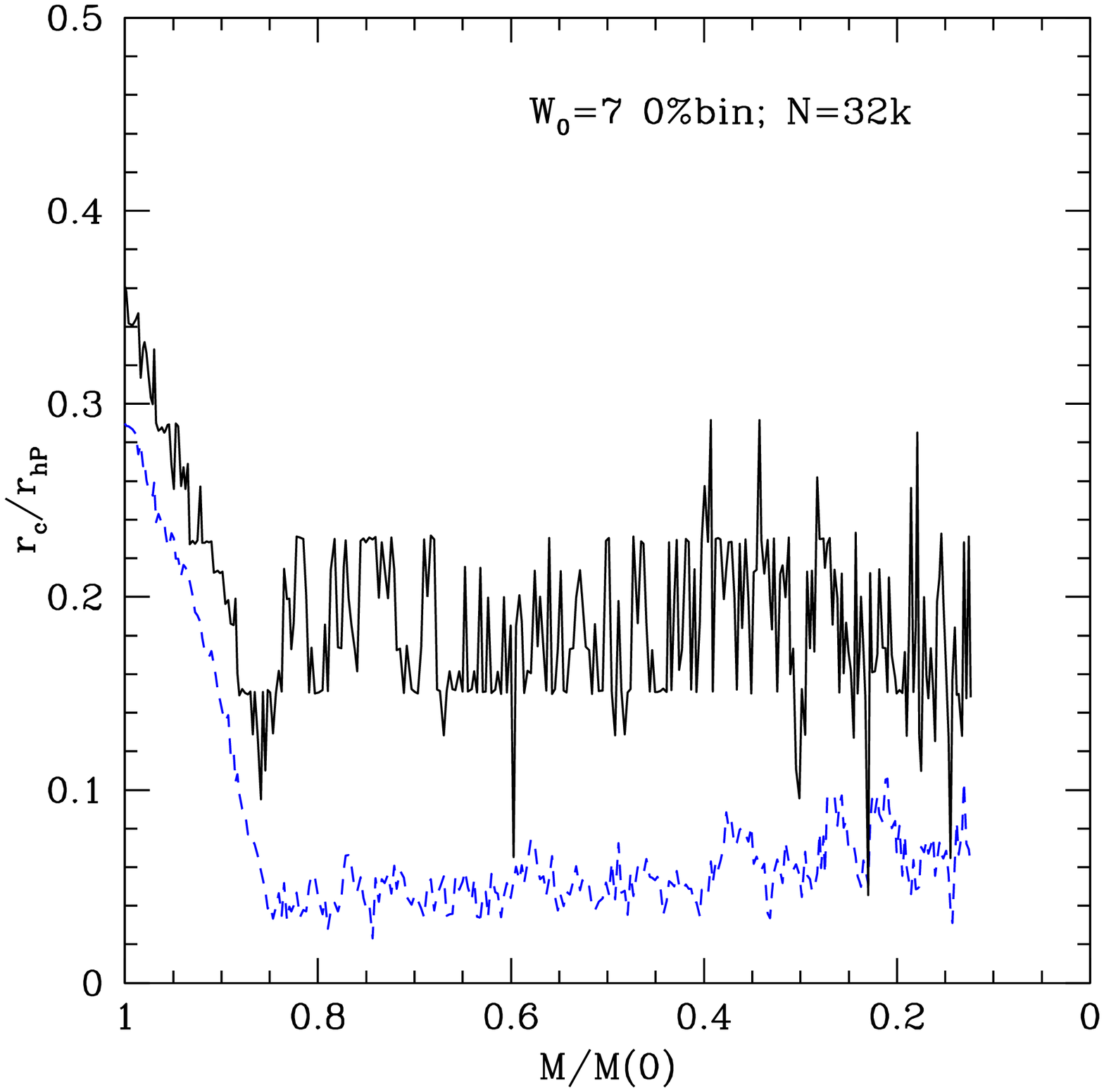}}
\resizebox{250pt}{!}{\includegraphics{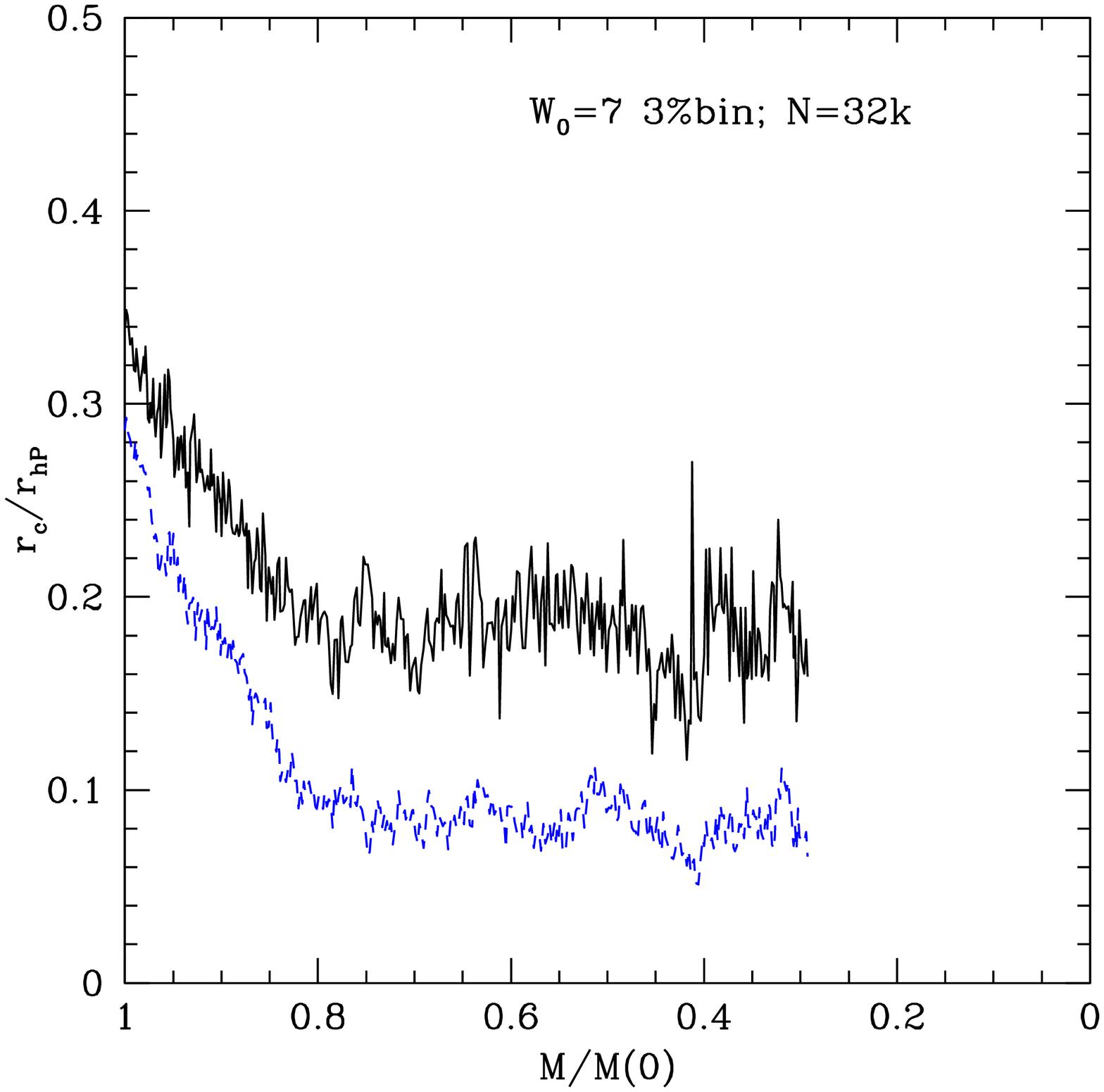}}
\resizebox{250pt}{!}{\includegraphics{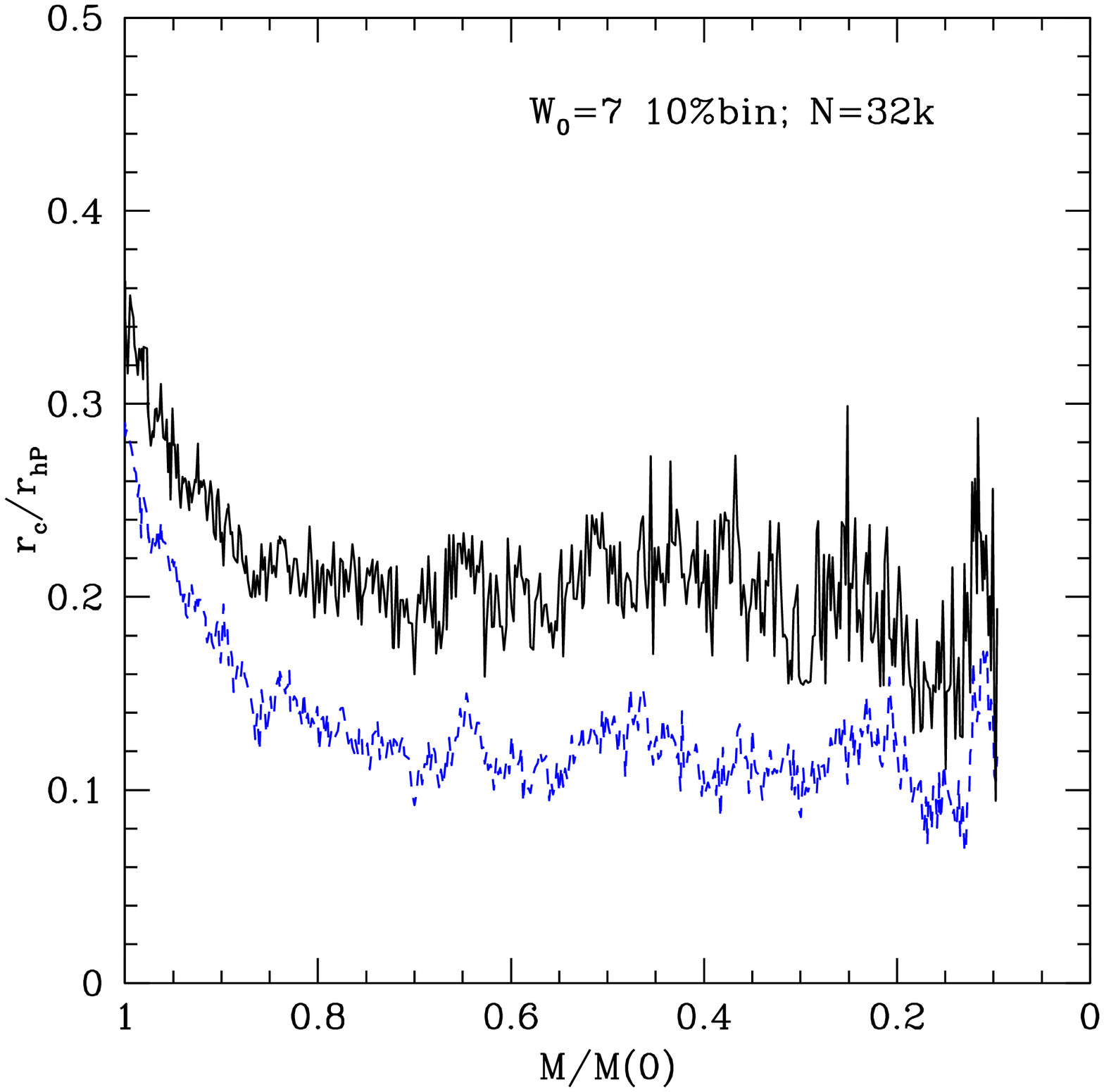}}

\caption{Observed core to half-light radius ratio ($r_c/r_{hP}$ solid
  black line) and 3D-density based core to half-mass radius ratio
  ($\sd{r_c/r_h}$ dashed blue line) for a $W_0=7$ model without
  primordial binaries (N=64k, upper left and N=32k, upper right) as
  well as with 3\% binaries (N=32k, bottom left) and with 10\% binaries (N=32k,
  bottom right). The two definitions differ significantly after core
  collapse, especially for models with a low fraction of primordial binaries. Core
  collapse signatures are clearly present in $\sd{r_c/r_h}$ for models
  without binaries but not in $r_c/r_{hP}$. }\label{fig:rcrh_comp}
\end{figure}

The initial density profile of the cluster or the tidal density field
is not important in the long term, as the memory of the starting
concentration is erased on a relaxation timescale, with a similar
subsequent evolution for models starting from different initial
conditions (see figs.~\ref{fig:alpha_m} to \ref{fig:c_m_64}).
$r_c/r_{hP}$ appears to be stable also over different IMFs, but its value
depends on the retention fraction of neutron stars and stellar mass
black holes. In particular a deep core collapse can be identified from
the core radius as determined following the observational procedures,
$r_c$, 
only if few dark remnants are present in the system. Even a
significant fraction (f=10\%) of primordial binaries does not appear
to influence much the observational determination of the core
radius. This is likely related to the heating induced 
by mass segregation of dark remnants in the core of the system.

\begin{figure} \resizebox{150pt}{!}{\includegraphics{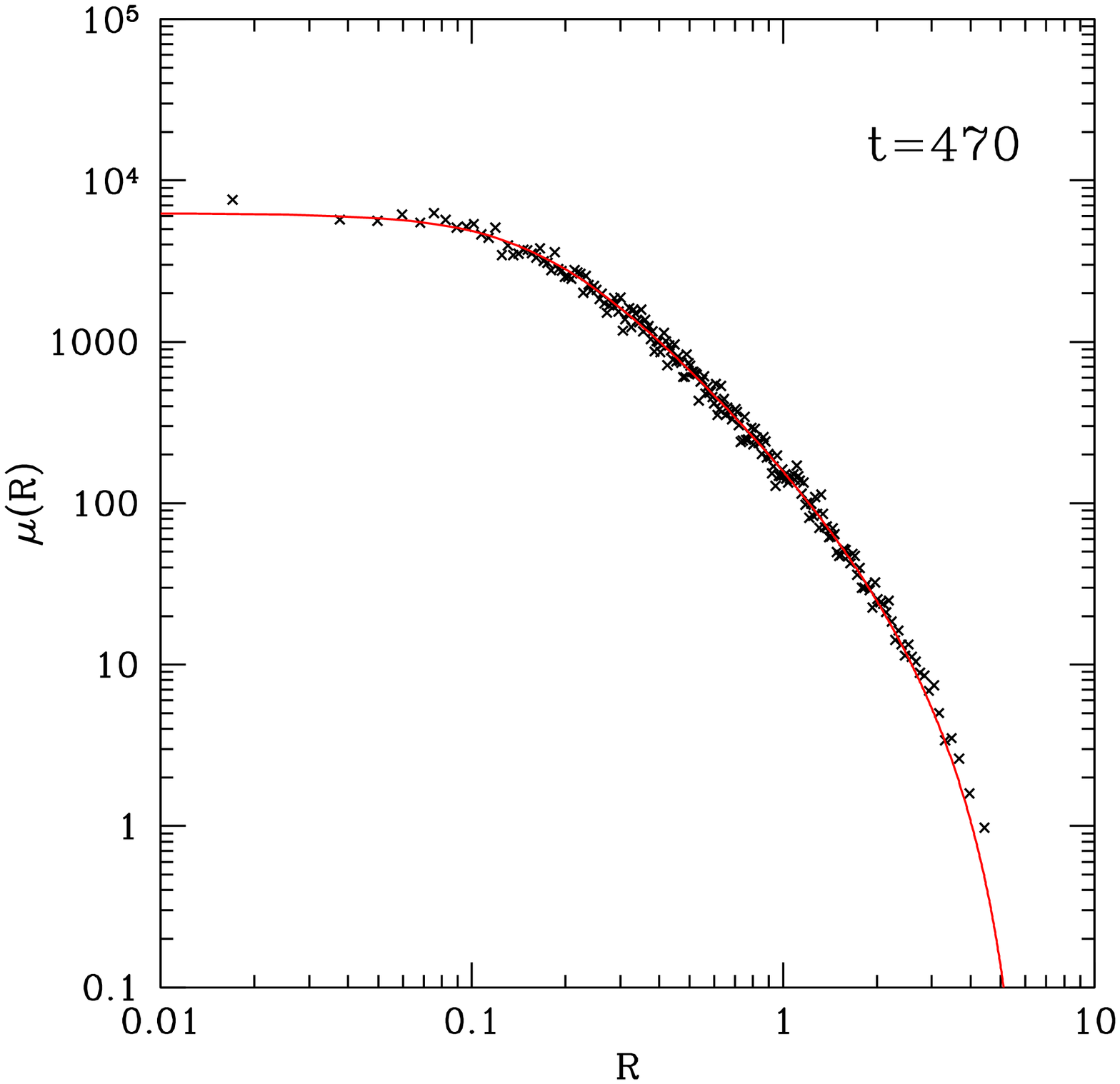}}
  \resizebox{150pt}{!}{\includegraphics{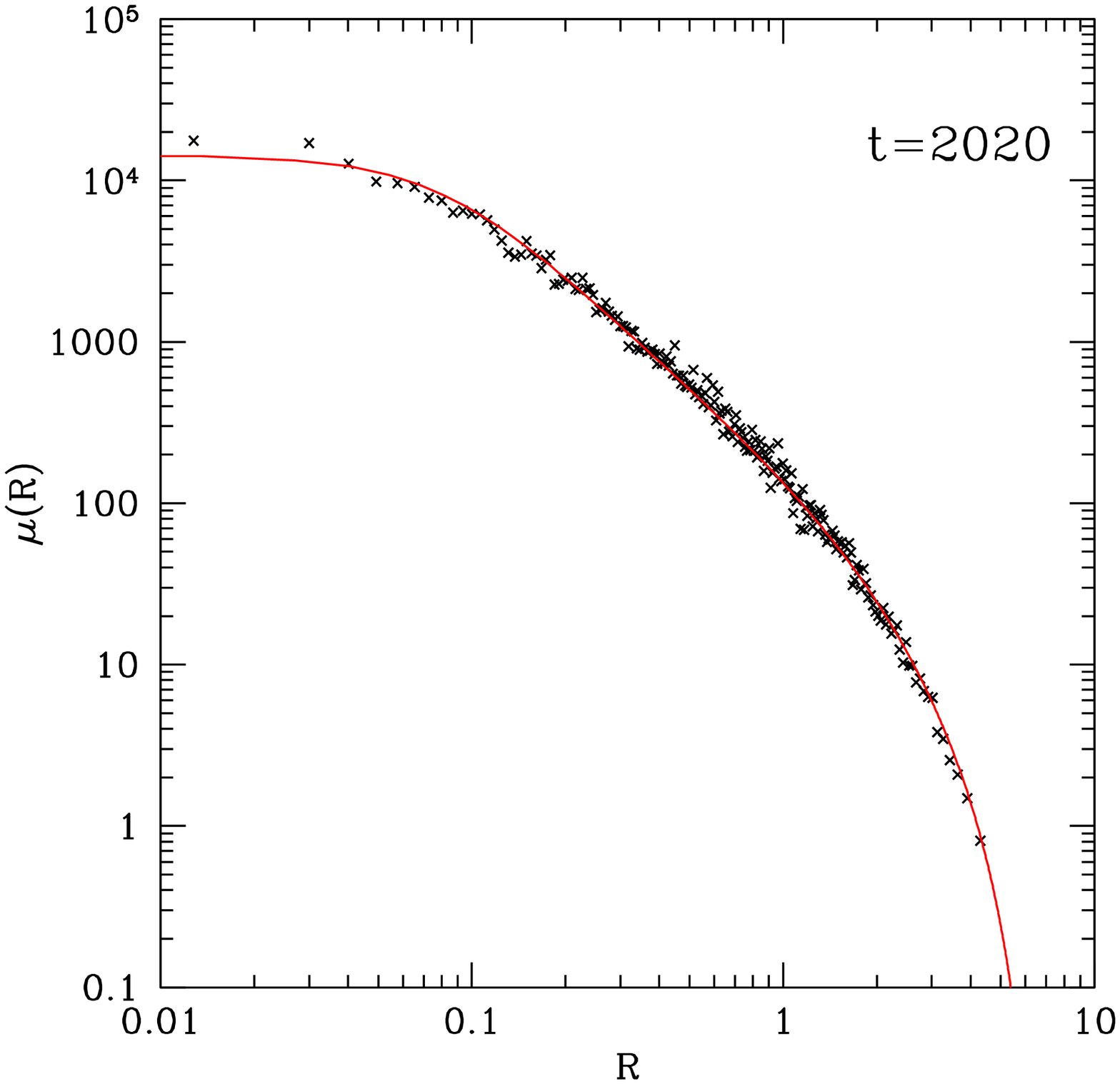}}
  \resizebox{150pt}{!}{\includegraphics{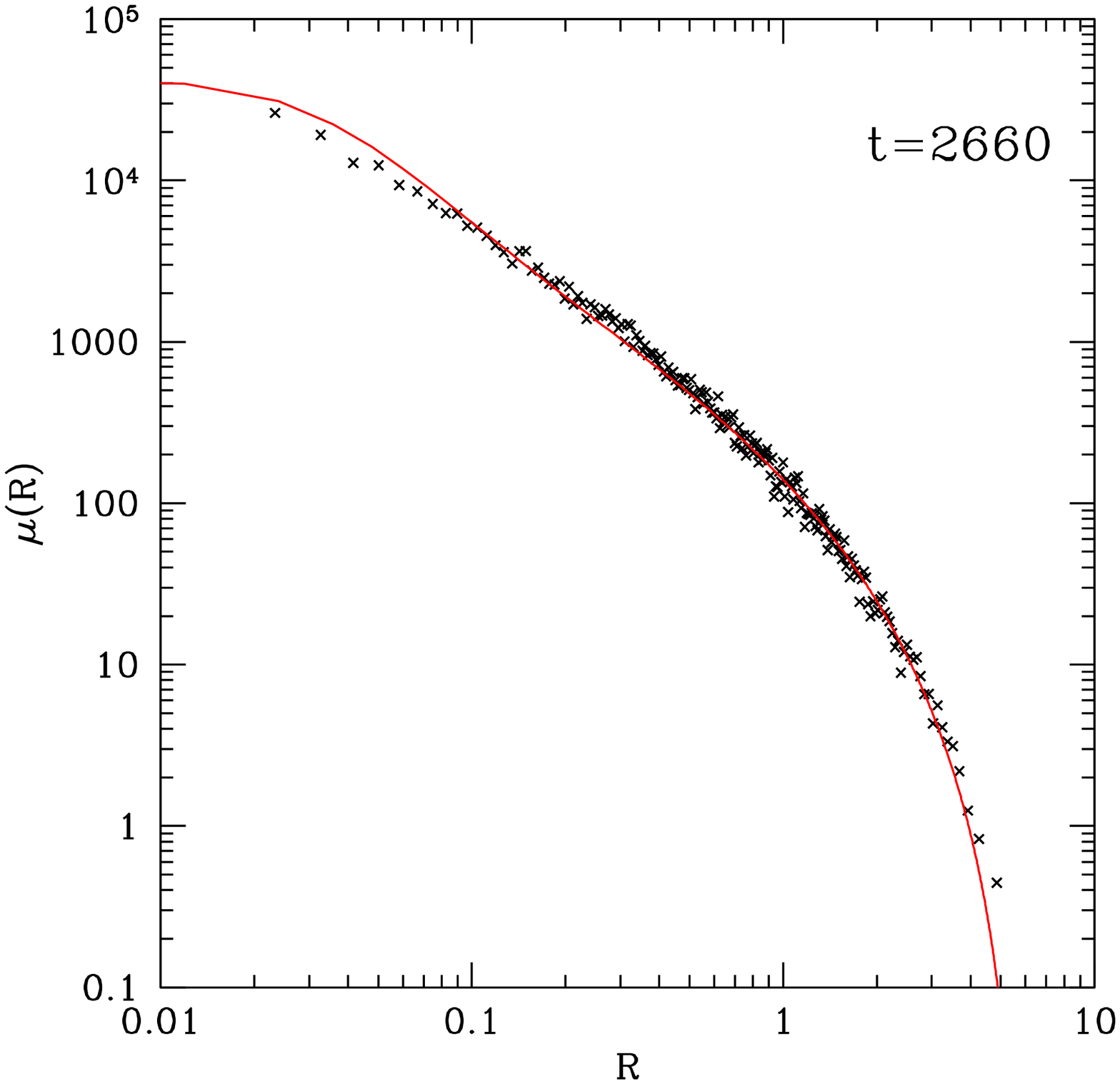}}
  \resizebox{150pt}{!}{\includegraphics{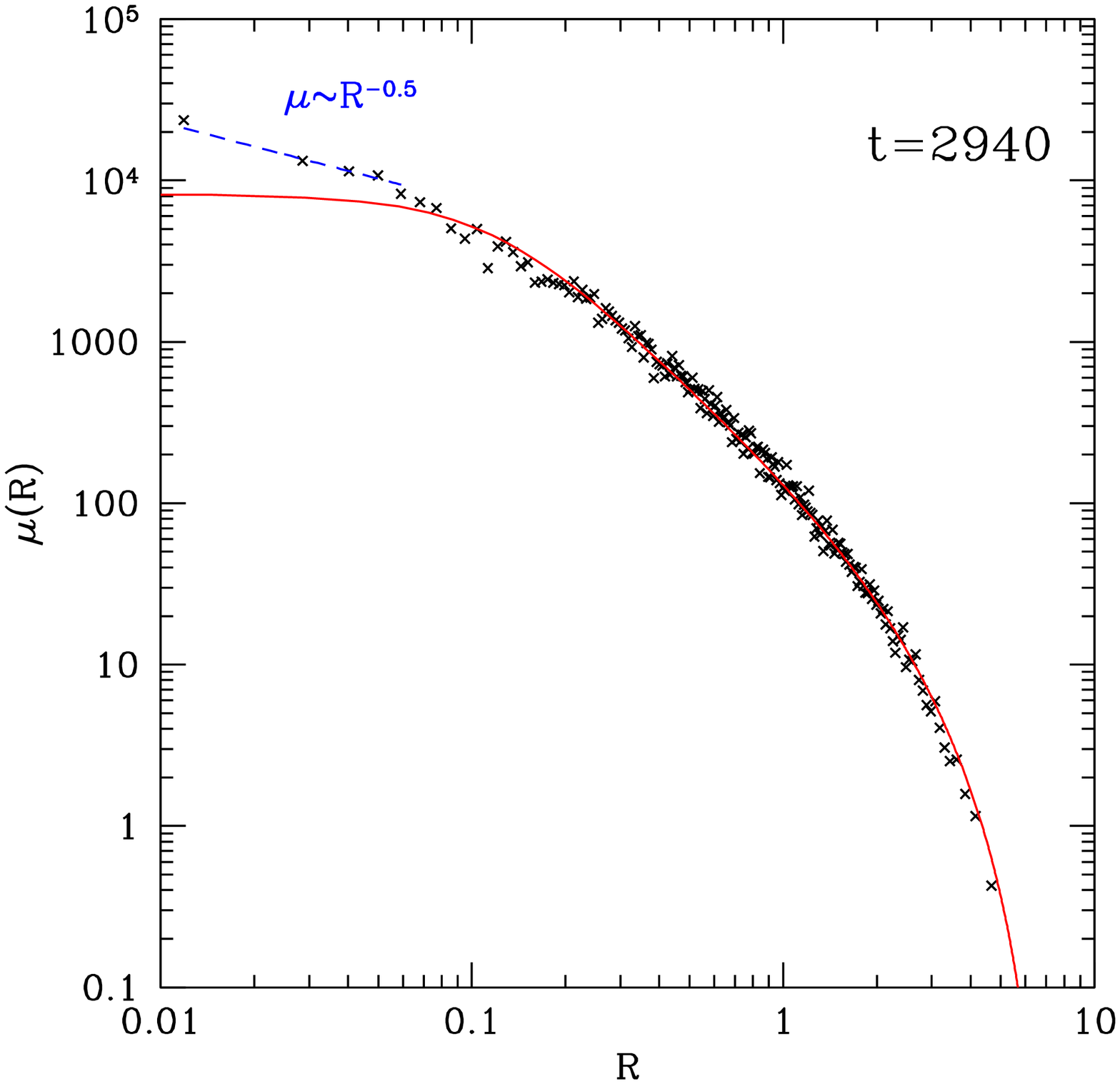}}
  \resizebox{150pt}{!}{\includegraphics{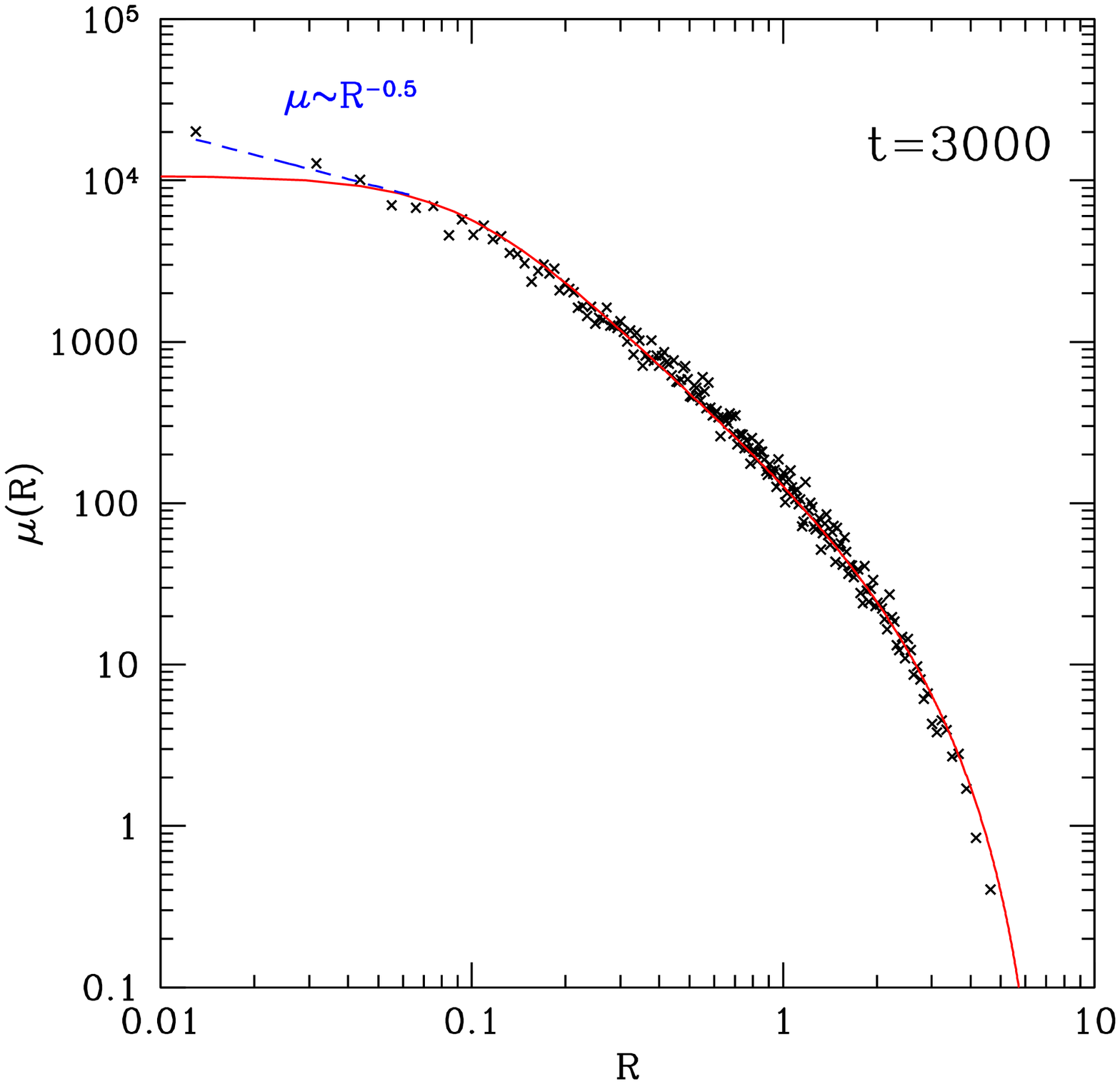}}
  \resizebox{150pt}{!}{\includegraphics{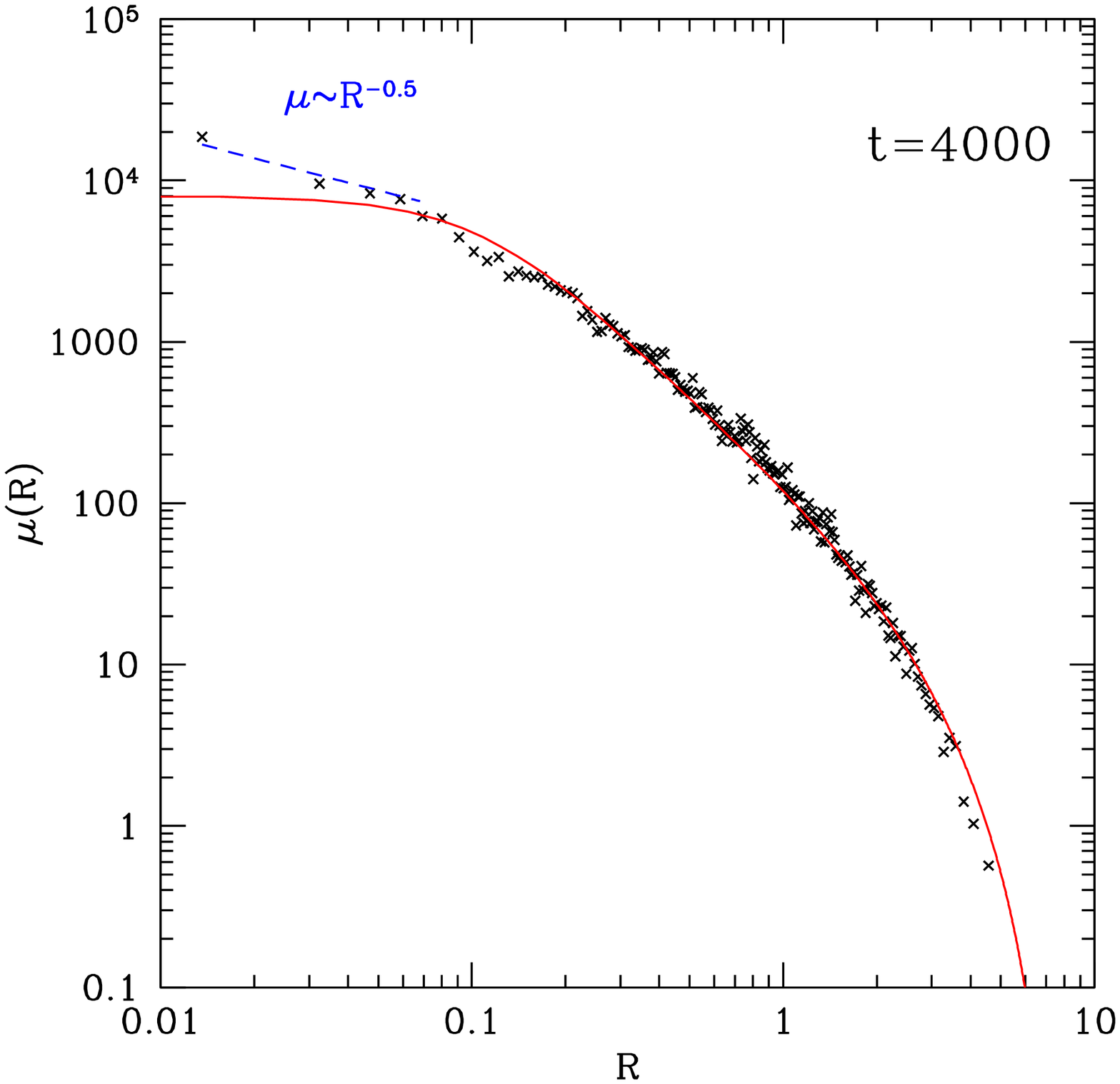}}
  \resizebox{150pt}{!}{\includegraphics{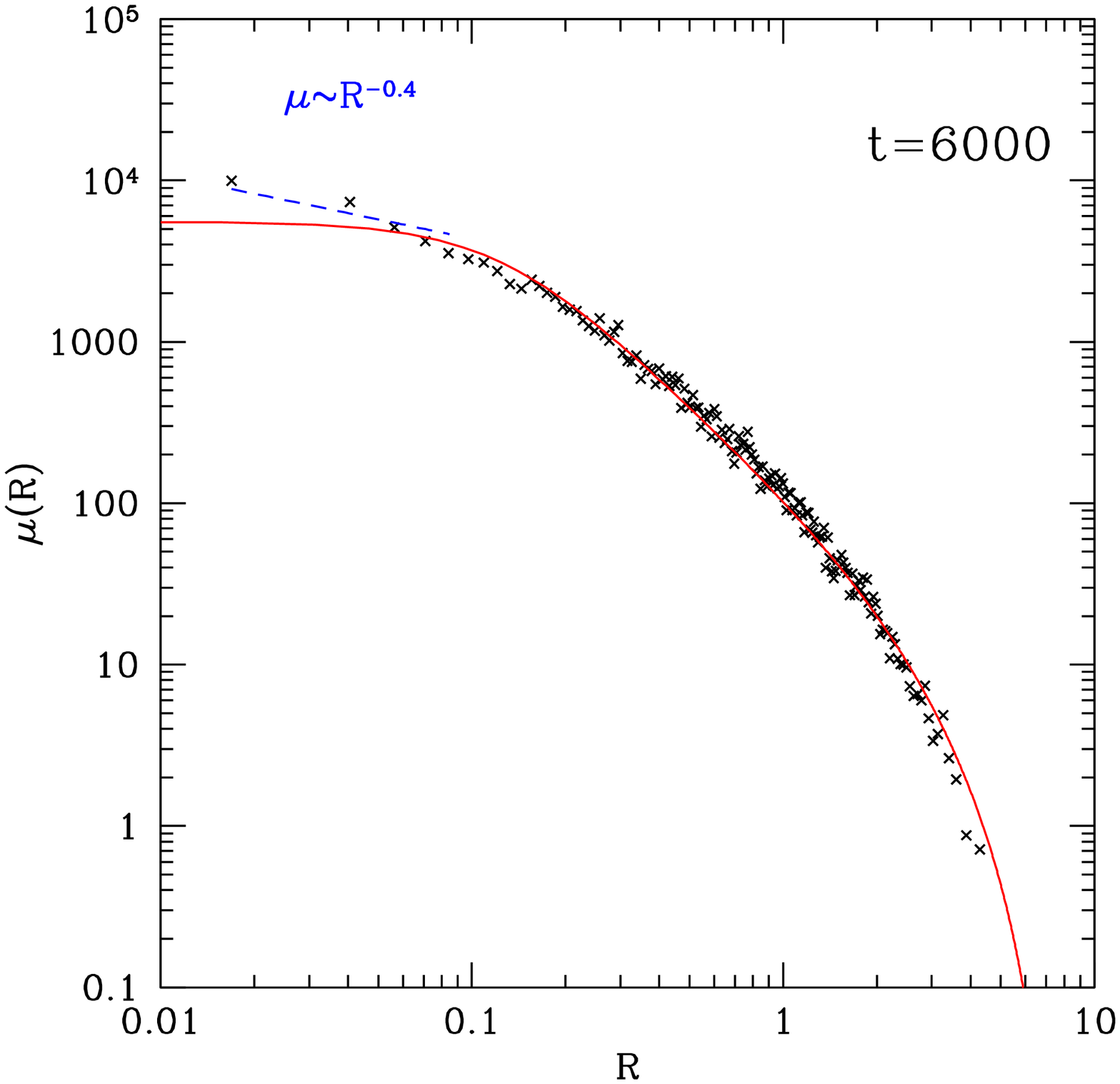}}
  \resizebox{150pt}{!}{\includegraphics{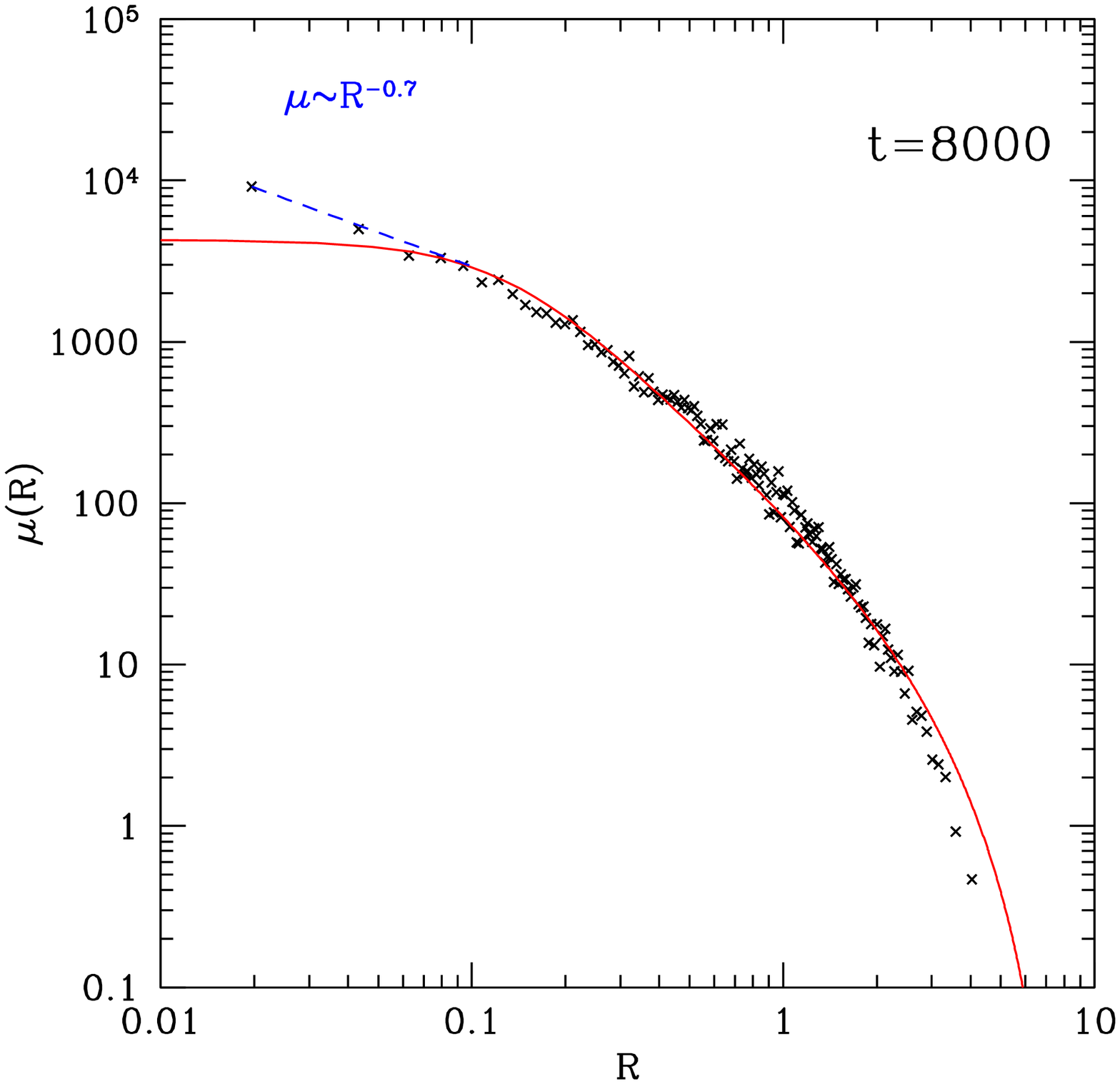}}
 
\caption{Surface
    brightness profiles (arbitrary units for $\mu$, NBODY units for
    $R$) for eight snapshots of our $N=64k$, $W_0=7$
    simulation with no binaries. The snapshots are taken at times
    $t=470,~2020,~2660,~2940,~3000,~4000,~6000,~8000$ in code units (equivalent to $\sim 1.0,~4.4,~5.8,~
    6.4, ~6.5,~8.7,~13.1,~17.4 ~ t_{rh}(0)$) and represents a typical pre-core-collapse
    profile, a profile on the way to core-collapse, the profile at the peak of core-collapse and five post-core-collapsed profiles. Superimposed to
    the data (black crosses) the best fit King model is shown. The
    five models have $W_0=7,~8.1,~9.4,~7.7,~8.0,~7.9,~7.7,~7.7$. The quality of the fit in the
    post-core-collapse profile is about two times worse than in the
    other snapshots. In the post-core collapse profiles (last five panels) a shallow
    central cusp ($\mu \sim R^{-\nu}$ with $\nu \sim 0.4-0.7$) is present (blue
    dashed line).}\label{fig:sb_fit}
\end{figure}

\begin{figure} 
\plotone{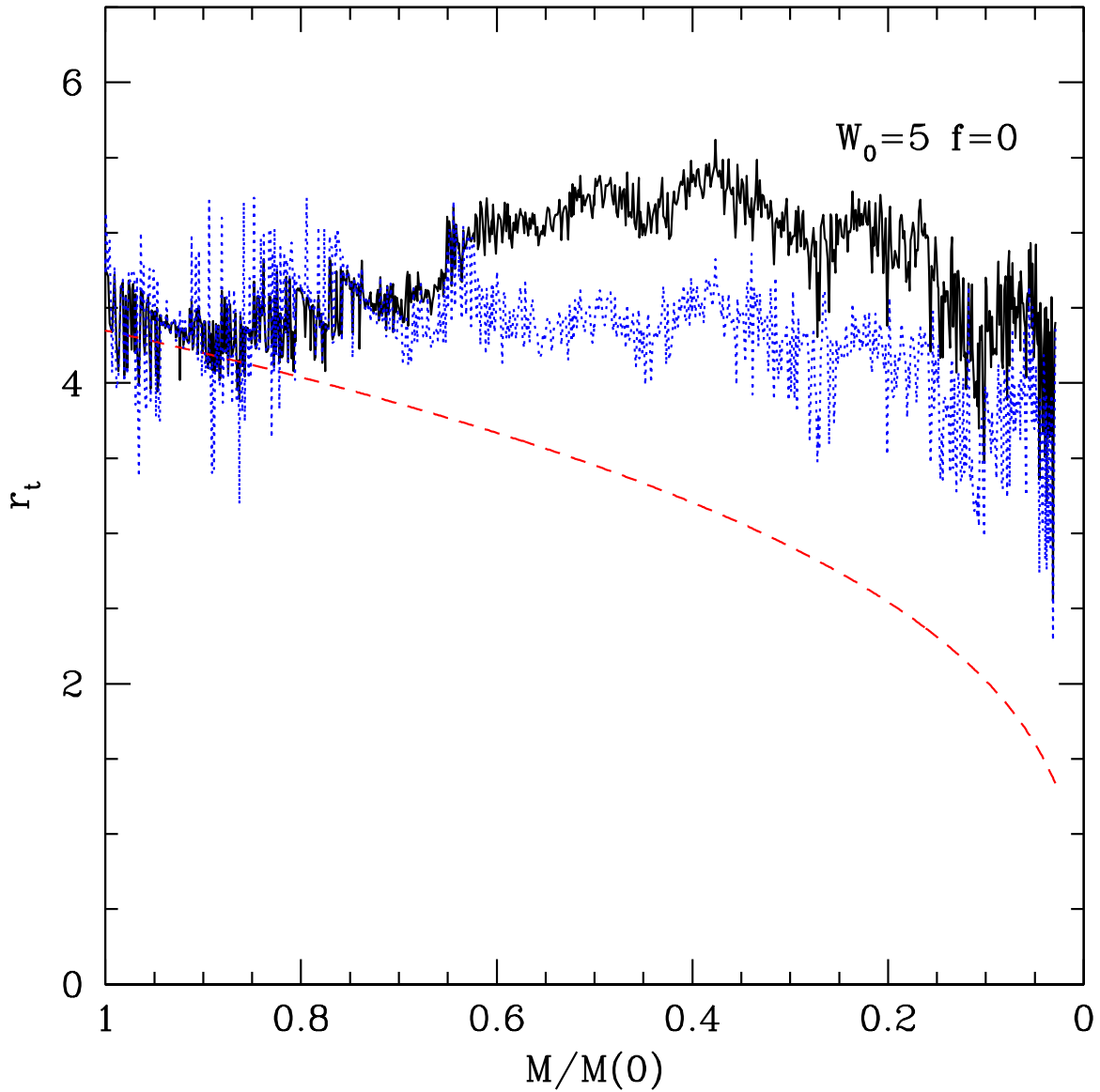}\caption{Tidal radius as measured
  observationally using the full surface brightness profile ($r_t$ --- solid black
line) or only the inner part containing 75\% of the total light (dotted blue line)
in our N=64k $W_0=5$ simulation
  with no primordial binaries. If information on the outer parts of the system is missing, 
then the tidal radius determination becomes dependent upon the details of the fitting method after core-collapse ($M/M(0)\lesssim 0.65$). The theoretical tidal radius is also shown
  ($\sd{r_t}$ --- red dashed line). At later times $r_t/\sd{r_t} \sim
  2$ when using the full surface brightness profile for the King
  model fit.}\label{fig:tidal}
\end{figure}

\begin{figure} \plotone{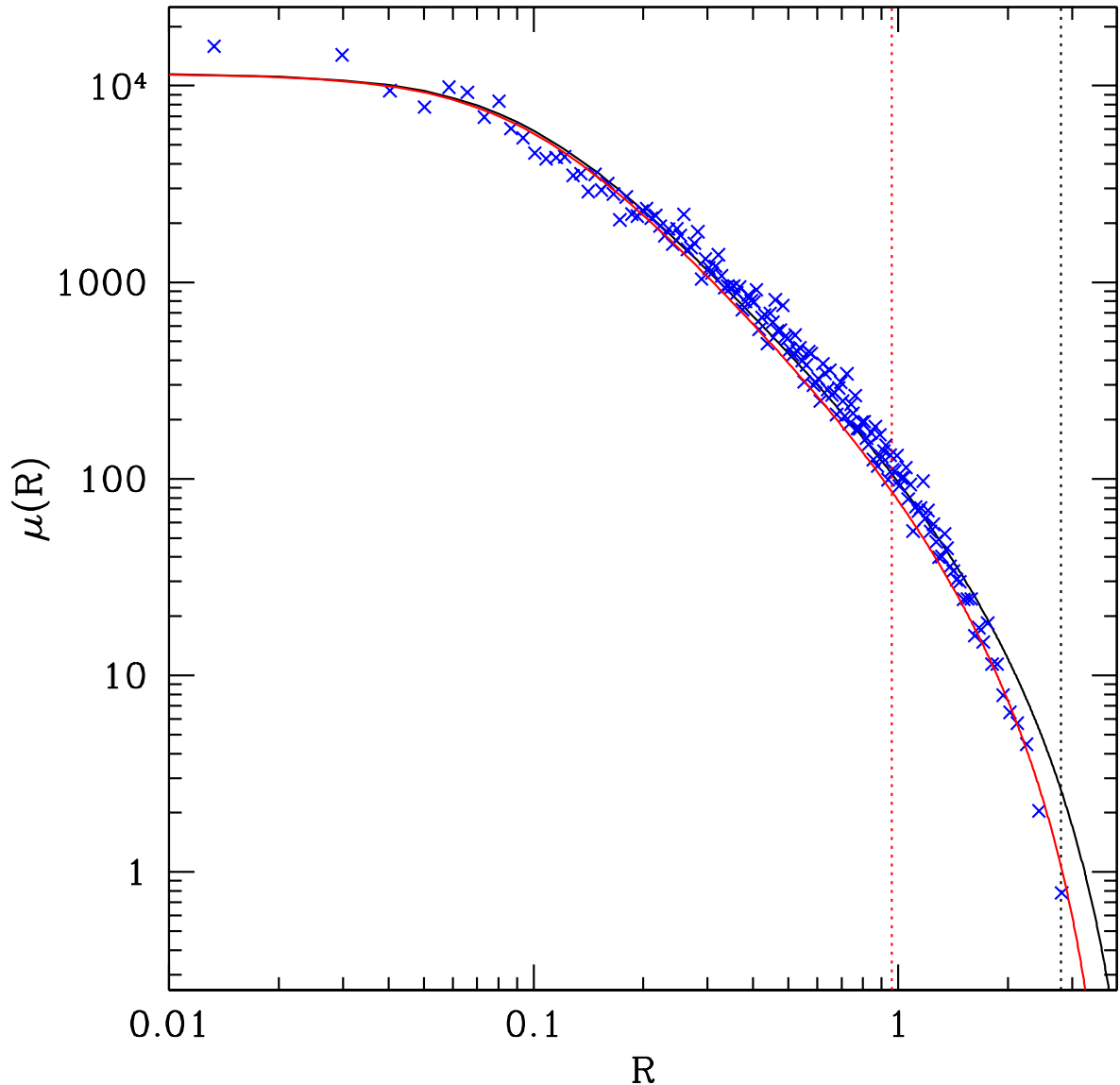}\caption{Surface brightness
    profile (blue crosses) at t=3420 (in the post-core collapse
    phase) for $N=64k$ simulation with no primordial binaries starting
    from $W_0=5$. The standard King model fit over the full radial range of the
    profile, and with total luminosity and half-light radius fixed to
    the measured quantities, is shown as black solid line ($W_0=7.5$). A King
    profile fit using only the inner 75\% of surface brightness
    profile points and with free total luminosity and half-light
    radius is shown as red solid line ($W_0=7.3$). The dotted vertical lines
    (black and red) show the outer boundary of the region used for
    fitting the King models in the first and second
    case.}\label{fig:tidal_fit} \end{figure}

\begin{figure} 
 \plottwo{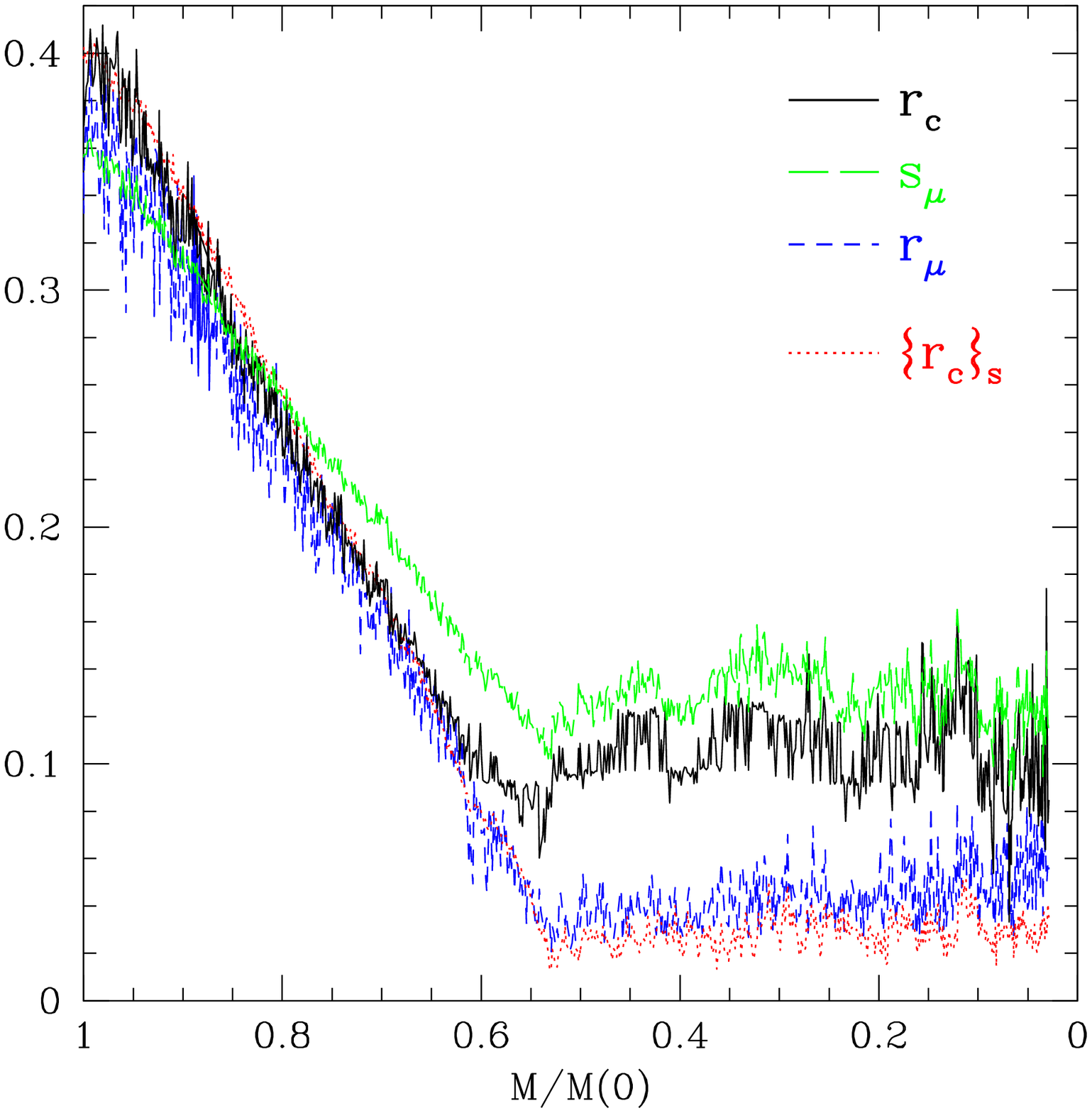}{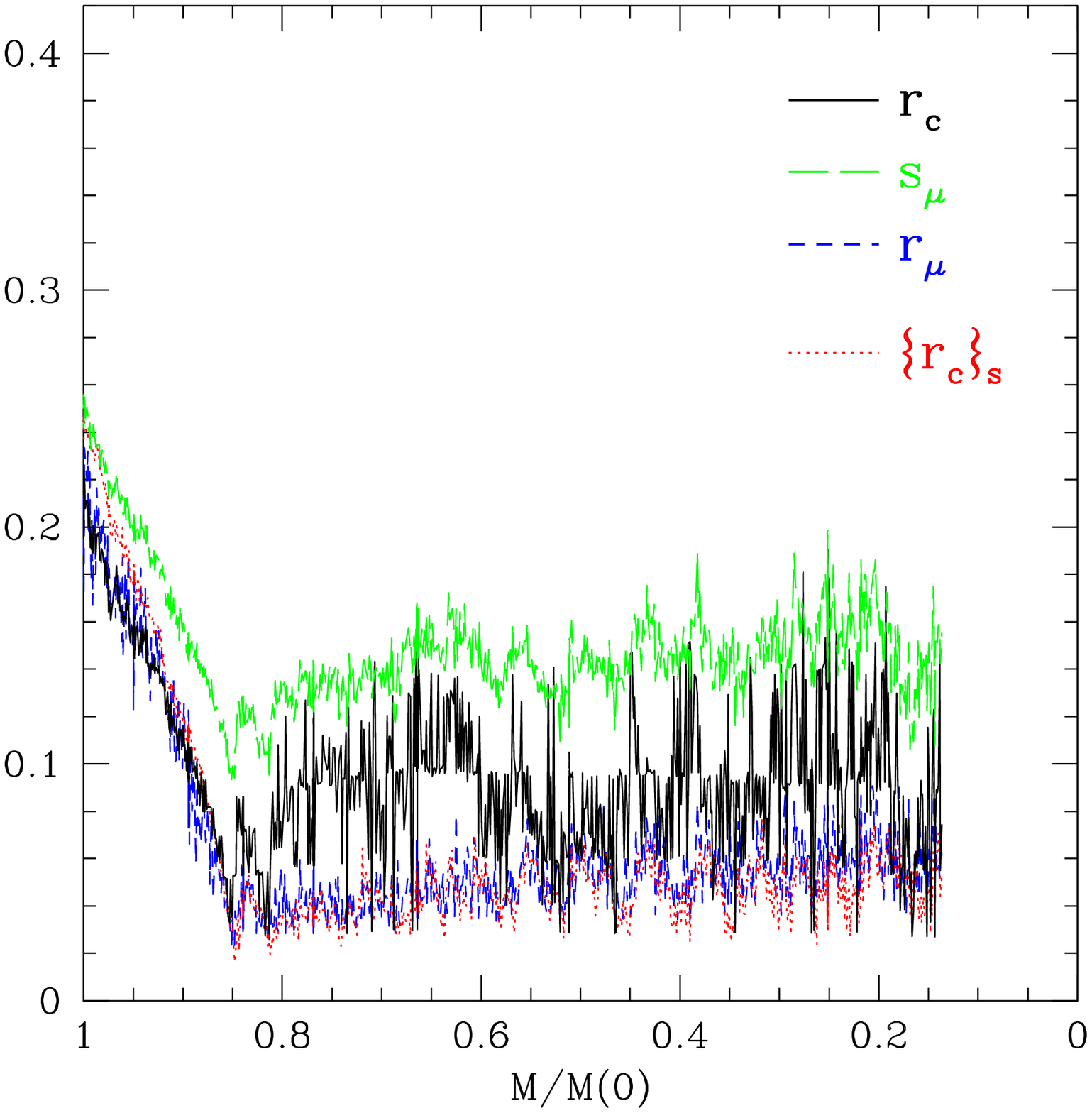} \caption{Comparison of
   different core radii definitions for two of our simulations without primordial binaries (left
   panel, N=64k, $W_0=5$; right panel N=32k, $W_0=7$, retention
   fraction 30\%). Post core-collapse oscillations in $r_c$ are
   particularly strong in the right panel.}\label{fig:rc_defs}
\end{figure}

\begin{figure} 
\plotone{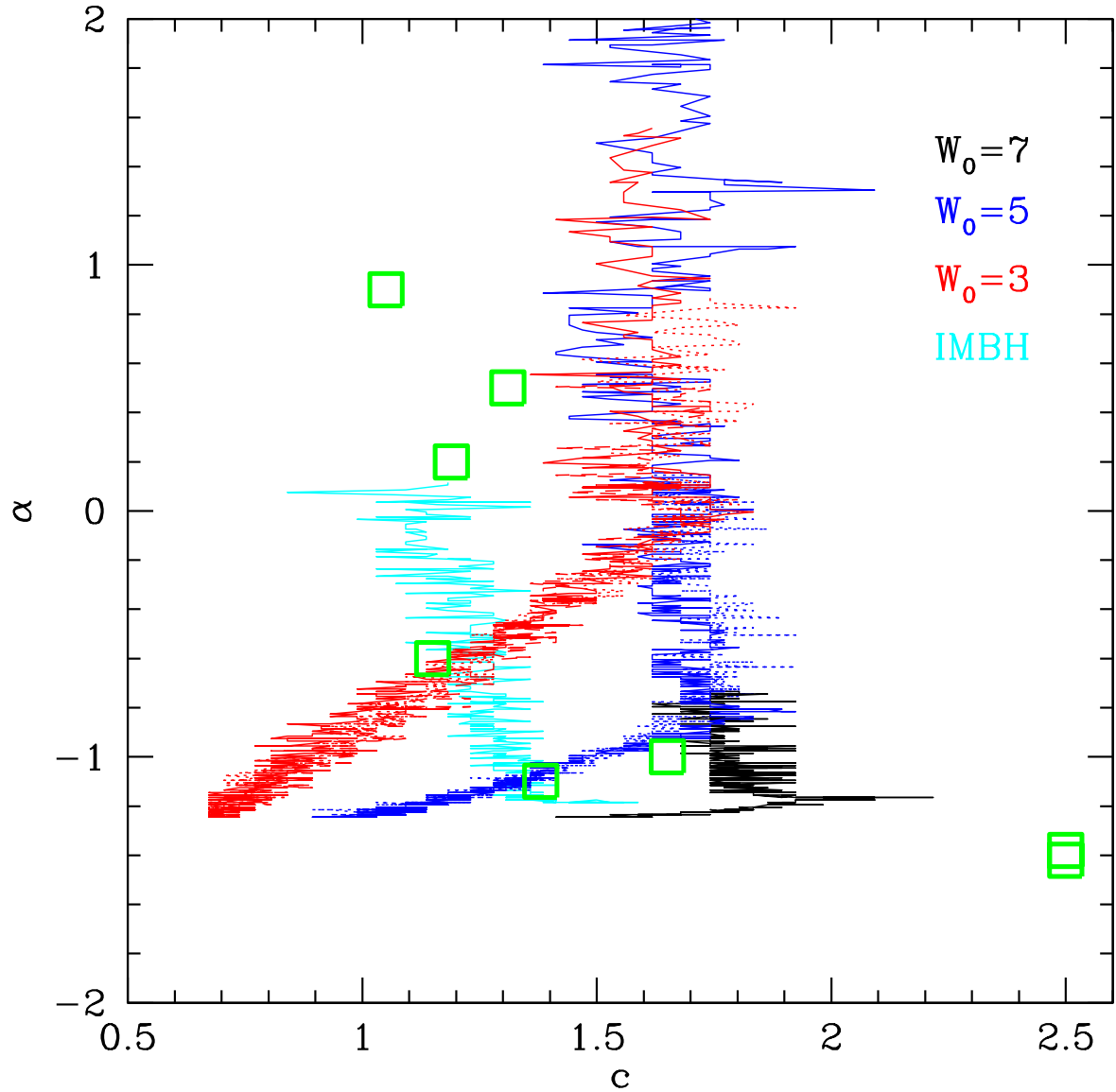}\caption{Mass function index $\alpha$ versus the
  central concentration for the $N=64k$ models of
  fig.~\ref{fig:alpha_m} and for the $N=32k$ model with a central IMBH (cyan solid line). The green squares are points from
  \citet{dem07} associated to globular clusters with $t_{rh} \leq 1$
  Gyr.}\label{fig:c_alpha}
\end{figure}

These findings highlight that the core radius $r_c$ as obtained from
King model fits over the complete radial extent of the surface
brightness profile are unable to reliably identify a core-collapse
cluster. Post-core collapsed systems are in fact relatively well
fitted by medium concentration King models, although high residuals
are present at the center of the system (see Fig.~\ref{fig:sb_fit}).
It is suggestive to note that these residuals appear similar to
shallow cusps that have been observed in several galactic globular
clusters with HST photometry \citep{noyola06}. In our simulations a
shallow cusp ($\mu \sim R^{-\nu}$ with $\nu \sim 0.4-0.7$) in the
surface brightness profile appears after core collapse (see
Fig.~\ref{fig:sb_fit}). The presence of such cusps may thus provide a
hint that a system is past core-collapse, but other indicators are
needed. One possibility if high angular resolution data are available,
is to use $r_{\mu}$ as a measure of the core, that is the radius at
which the surface brightness falls to half of its central value.
$r_{\mu}$ does in fact track the evolution of the tri-dimensional core
radius defined in simulations, i.e. $\sd{r_c}$ (see
Fig.~\ref{fig:rc_defs}). In alternative, one could explore the use of
the mass-to-light ratio if spectroscopic information are available.
Shallow cusps at the center of medium concentration surface brightness
profiles appears to be so common in the post core-collapsed evolution
of a globular cluster that this indicator is not particularly helpful
in suggesting clusters likely to harbor a central IMBH, contrary to
past suggestions \citep{baum05,miocchi07}.

Our investigation on the evolution of structural parameters
``observed'' from simulations shows an improved agreement with the
$c-\alpha$ correlation identified by \citet{dem07} for Galactic
globular clusters, but still the simulated clusters are too
concentrated in their final stages of their evolution. This might be
related to the definition of the tidal radius $r_t$, and thus of the
concentration, from the $\gtrsim 15$ years old, ground-based data used
by \citet{trager95}. The light profile is in fact dominated by
contributions from stars along the giant branch, more massive than
average, and therefore more mass-segregated than the main-sequence
stars we used in this analysis. In any case, from the observed
$c-\alpha$ correlation we can expect, based on our analysis, that some
of the systems with the most depleted stellar mass function can indeed
be core-collapsed systems lurking in the sample of objects fitted by
regular King profiles. This highlights the importance of carrying out
a uniform survey to construct the structural parameters of Milky Way
globular clusters, combining large area coverage with the high
angular-resolution data at the center, like those obtained for about
50 globulars by \citet{sara07} with HST.

\acknowledgements

We thank Douglas Heggie and the referee for useful suggestions. We are
grateful to Sverre Aarseth for his dedication to constantly improving
its NBODY6 code and for making every improvement publicly available.
We also thank Keigo Nitadori for the development of the CUDA/OpenMP
extension of the code. MT and EV acknowledge partial support from NASA
ATFP grants NNX08AH29G, NNX08AH15G and from grant HST-AR-11284,
provided by NASA through a grant from STScI, which is operated by
AURA, Inc., under NASA contract NAS 5-26555. All the authors
acknowledge support from the Kavli Institute for Theoretical Physics,
through the National Science Foundation grant PHY05-51164. This
research was supported in part by the National Science Foundation
through TeraGrid resources provided by the National Center for
Supercomputing Applications (grants TG-AST090045 and TG-AST090094).

%%%%%%%%%%%%%%%%%%%%%%%%%%%%%%%%% 

%%%%%%%%%%%%%%%%%%%%%%%%%%%%%%%%%%%%%%%%%%%%%%%%%%%%%%%5
\clearpage

\clearpage

\clearpage

\clearpage

\clearpage

\clearpage

%%%%%%%%%%%%%%%%%%%%%%%%%%%%%%%%%%%%%%%%%%%%%
%%%%%%%%%%%%%%%%%%%%%%%%%%%%%%%%%%%%%%%%%%%%%%%%
\clearpage

\clearpage

%%%%%%%%%%%%%%%%%%%%%%%%%%%%%%%%%%%%%%%%%%%%%
%%%%%%%%%%%%%%%%%%%%%%%%%%%%%%%%%%%%%%%%%%%%%%%%Giersz
\clearpage

%%%%%%%%%%%%%%%%%%%%%%%%%%%%%%%%%%%%%%%%%%%%%
%%%%%%%%%%%%%%%%%%%%%%%%%%%%%%%%%%%%%%%%%%%%%%%%Giersz
\clearpage

%%%%%%%%%%%%%%%%%%%%%%%%%%%%%%%%%%%%%%%%%%%%%
%%%%%%%%%%%%%%%%%%%%%%%%%%%%%%%%%%%%%%%%%%%%%%%%
\clearpage

%%%%%%%%%%%%%%%%%%%%%%%%%%%%%%%%%%%%%%%%%%%%%
%%%%%%%%%%%%%%%%%%%%%%%%%%%%%%%%%%%%%%%%%%%%%%%%Giersz
\clearpage

%%%%%%%%%%%%%%%%%%%%%%%%%%%%%%%%%%%%%%%%%%%%%%%%%%
\begin{table}
\begin{center}
\caption{Summary of N-body simulations\label{tab:sim}}
\begin{tabular}{ccccc}
\tableline\tableline
$N$ & $f$ & IMF & $W_0$ & Other \\ 
65536 & 0.00& MS & 3 &\\
65536 & 0.02& MS & 3 &\\
65536 & 0.055& MS & 3 &\\
65536 & 0.00& MS & 5 &\\
65536 & 0.02& MS & 5 &\\
%65536 & 0.05& MS & 5 &\\
65536 & 0.00& MS & 7 &\\
%65536 & 0.02& MS & 7 &\\
%65536 & 0.05& MS & 7 &\\
32768 & 0.00& MS & 7 &\\
32768 & 0.01& MS & 7 &\\
32768 & 0.03& MS & 7 &\\
32768 & 0.05& MS & 7 &\\
32768 & 0.10& MS & 7 &\\
32768 & 0.00& Sa & 7 &\\
32768 & 0.00& MS & 7 & 30\% NS/BH retention\\
32768 & 0.00& Sa & 7 & 30\% NS/BH retention\\
32769 & 0.00& MS & 7 & $m_{IMBH}=0.01$\\
32769 & 0.00& MS & 3 & $\sd{r_t}=6.28$ [underfilled 2]\\
\tableline 
\end{tabular}
%% Any table notes must follow the \end{tabular} command.
\tablecomments{N-body simulations of star clusters in a tidal field
  with self consistent King model initial conditions. First column:
  number of particles $N$; second column:
  binary fraction $f=N_b/N$; third column: initial mass function used
  (Sa: power law, MS: Miller \& Scalo); fourth column: initial
  concentration of the density profile (King index $W_0$). Fifth
  column reports other relevant information for the simulation. The
  last simulation starts from compact initial conditions where the
  Roche lobe is underfilled by a factor 2. }
 %%% \tablenotetext{a}{ ~Low tidal field intensity.}
\end{center}
\end{table}

\end{document}